# Hydrogen Trapping and Embrittlement in Metals – A Review

Yi-Sheng Chen[1,2,3,*], Chao Huang[1,2], Pang-Yu Liu[1,2], Hung-Wei Yen[3,4], Ranming Niu[1,2], Patrick Burr[5], Katie L. Moore[6,7], Emilio Martínez-Pañeda[8], Andrej Atrens[9], Julie M. Cairney[1,2,*]

[1] Australian Centre for Microscopy and Microanalysis, The University of Sydney, Australia
[2] School of Aerospace, Mechanical and Mechatronic Engineering, The University of Sydney, Australia
[3] Department of Materials Science and Engineering, National Taiwan University, Taiwan
[4] Advanced Research Center For Green Materials Science and Technology, National Taiwan University, Taiwan
[5] School Mechanical and Manufacturing Engineering, University of New South Wales, Australia
[6] Department of Materials, The University of Manchester, United Kingdom
[7] Photon Science Institute, The University of Manchester, United Kingdom
[8] Department of Civil and Environmental Engineering, Imperial College London, United Kingdom
[9] School of Mechanical and Mining Engineering, The University of Queensland, St Lucia, 4072, Australia

*Corresponding authors: yi-sheng.chen@sydney.edu.au; julie.cairney@sydney.edu.au

## Abstract

Hydrogen embrittlement in metals (HE) is a serious challenge for the use of high strength materials in engineering practice and a major barrier to the use of hydrogen for global decarbonization. Here we describe the factors and variables that determine HE susceptibility and provide an overview of the latest understanding of HE mechanisms. We discuss hydrogen uptake and how it can be managed. We summarize hydrogen trapping and the techniques used for its characterization. We also review literature that argues that hydrogen trapping can be used to decrease HE susceptibility. We discuss the future research that is required to advance the understanding of HE and hydrogen trapping and to develop HE-resistant alloys.

## Table of content

**1. Introduction**

Hydrogen (H), the simplest and the most abundant element in the universe, is widely used in industry for chemical processing, including oil refining and the production of ammonia and methanol. It can also be used be used to store, transport, and deliver energy. Recent years have seen a dramatic increase in its use in emerging energy technologies.

The hydrogen molecule ($H_2$) has an energy density of 120 MJ/kg, which is 3 times higher than that of gasoline and is the highest among all known fuels [1]. Hydrogen electrolysis and hydrogen fuel cells consume and produce electricity on demand, potentially alleviating load-demand issues in intermittent renewable energy grids. Hydrogen fuel can be used for combustion and electrification and is free of intrinsic carbon emission. It can hence facilitate the decarbonization of the transportation sector, which currently depends heavily on fossil fuels. Hydrogen can also serve as a reductant or fuel to decarbonize hard-to-abate high-emission sectors such as steelmaking [2–4]. It is possible to produce hydrogen at scale from a variety of sources, including natural gas, coal, biomass, waste plastics, water electrolysis [5], and nuclear power [6]. Production methods that have a carbon footprint may be combined with carbon capture and storage technologies to reduce the embodied carbon emissions of hydrogen production [1]. Hydrogen is being increasingly considered as an essential commodity to replace fossil fuels [1,5]. Worldwide, national and international initiatives are underway to develop a 'hydrogen economy', including in the USA [5,7], Japan [8], UK [9], Germany [10], Norway [11], Canada [12], China [13], and Australia [14]. However, challenges must be addressed to enable this vision. Hydrogen is flammable in air at a concentration above the lower explosive limit [15] and requires careful handling [1]. Moreover, hydrogen can degrade metallic materials, reducing their fracture toughness, fatigue resistance and ductility. This effect is called hydrogen embrittlement (HE).

HE is one of the biggest obstacles for the deployment of hydrogen energy infrastructure, including the repurposing of natural gas pipelines for the transport of gaseous hydrogen [5,9,12,14,16–19]. Failure of energy infrastructure due to HE could lead to life-threatening accidents – just one major incident could cause a setback in the worldwide application of hydrogen for decarbonization [20–22]. In fact, a recent Australian national survey testing public acceptance of hydrogen energy indicated that safety is the most important factor in determining people's willingness to use hydrogen [23]. On a 5-point scale where 5 signifies the highest level of importance, 'safety' received an average rating of 4.5, higher than reliability (4.3), cost (4.2), and convenience (3.6). Beyond the energy sector, HE is an ongoing issue for the use of high-strength materials in defense, transport, and construction [24–26].

Although HE is a longstanding industrial problem, in recent years there has been potential mitigation approaches being proposed and verified based on microstructural hydrogen trapping, the advances of which were enabled through the improved characterization and modelling tools. Here we provide a contemporary overview of HE, summarize the latest efforts in using hydrogen trapping for mitigating HE, and outline promising future research directions.

**2. Hydrogen embrittlement**

Hydrogen can embrittle metals and alloys, a phenomenon that was first reported almost 150 years ago. In 1874, Johnson [27] found significant decreases in the breaking strain of steel and iron specimens after immersion in hydrogen-bearing solutions. This mechanical degradation was recovered after the hydrogen was fully desorbed, demonstrating that this degradation was a result of hydrogen uptake. Johnson also found that HE in higher-strength specimens was more significant than for those with similar compositions but with lower strength. Since this report, over 38,000 papers have been published on the subject [28], reflecting the engineering importance, multidisciplinary nature, and complexity of HE. Some previous review articles have provided a description of the HE phenomena and some understanding of the mechanistic causes [16,26,29–59].

Before going into detail, it is necessary to distinguish between internal hydrogen embrittlement (IHE) and environmental hydrogen embrittlement (EHE). IHE refers to hydrogen-induced failure caused by the presence of pre-existing hydrogen in the alloy and is typically limited by hydrogen supply. EHE is the response of the material to hydrogen when a specimen is subject to a mechanical load with simultaneous hydrogen charging. For EHE, the extent of hydrogen uptake depends on the charging time and the hydrogen diffusivity within the material. EHE failure typically initiates at the specimen surface, where environmental effects are most severe.

## 2.1. Variables and metrics

### 2.1.1. Hydrogen content

HE only occurs when a specimen is subjected to a sufficiently high stress (either applied or residual) and contains a hydrogen content above a critical level, as defined in **Figure 1**. The exact critical stress and hydrogen content depend on the type of alloy. Metals that require energy to absorb hydrogen (i.e., endothermic hydrogen solution), such as Fe and Ni, have low hydrogen solubility at ambient temperatures and pressures. These metals generally have a low critical hydrogen content, as low as parts per million in weight (wt. ppm, wppm, or µg/g). **Figure 1**A shows the ultimate tensile strength (UTS) of a martensitic steel as a function of hydrogen content. Just 5 wt. ppm of hydrogen significantly reduces the UTS, but less than 2 wt. ppm has little influence [60]. The hydrogen content in this figure was measured by using hydrogen thermal desorption analysis (TDA) [61] and the notched UTS was determined using a slow strain rate tensile test (SSRT) on ex-situ electrolytically hydrogen-charged specimens [45]. **Figure 1**A shows that hydrogen-induced fracture for hydrogen concentrations above 7 wt. ppm occurred at approximately 30% of the UTS in the absence of hydrogen. **Figure 1**B shows the threshold stress intensity factor for hydrogen assisted cracking ($K_{TH}$) versus diffusible hydrogen content for a martensitic steel (AerMet 100) [62], showing decreased $K_{TH}$ with increasing hydrogen content.

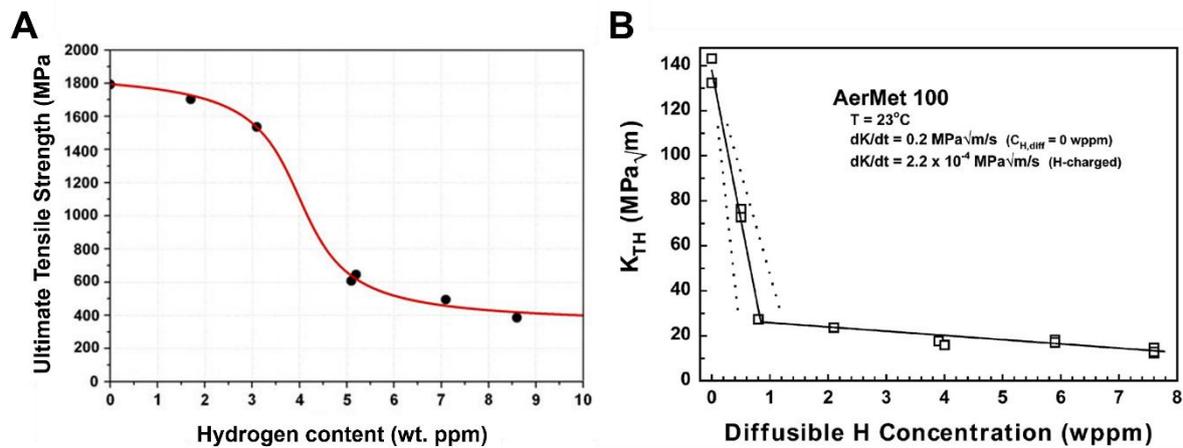

**Figure 1. Hydrogen content in common metals and HE susceptibility.** (A) The material notch strength (defined by the ultimate tensile strength) of a hydrogen-charged untempered martensitic steel decreased with increasing hydrogen content, reaching low values for hydrogen contents above 5 wt. ppm, reproduced from [60]. (B) The threshold stress intensity factor ($K_{TH}$) decreased with increasing diffusible hydrogen content in a AerMet 100 martensitic steel in the near peak aged condition with a nominal 1765 MPa yield strength. Hydrogen was charged with increasing cathodic potential to increase the hydrogen concentration. Reproduced from [62]

The service environment determines the hydrogen content in a material, and its measured HE susceptibility may be low if that environment causes a hydrogen content that is below the critical value for that material. For example, the advanced high-strength steel shown in **Figure 1**A is unlikely to experience HE in automotive applications because the hydrogen concentration in service is generally lower than 1 wt. ppm [63]. Note also that hydrogen uptake is sensitive to the mechanical stress state, as the solubility depends on the hydrostatic stress [64,65].

*2.1.2. Mechanical Properties*

HE can reduce alloy strength and/or ductility in tensile tests, as shown for two different steels in **Figure 2**A and B. Common metrics for HE susceptibility for tensile type specimens are strength and ductility loss:

$$Strength\ loss\ index = \frac{\sigma_{ref} - \sigma_H}{\sigma_{ref}} \quad (1)$$

$$Ductility\ loss\ index = \frac{RA_{ref} - RA_H}{RA_{ref}} \quad (2)$$

where $\sigma_H$ (or $RA_H$) is the strength (or reduction in area) in hydrogen and $\sigma_{ref}$ ($RA_{ref}$) is the strength (or reduction in area) in a hydrogen-free condition, typically air or an inert gas. Increased susceptibility to HE is manifested by a lower strength $\sigma_H$ or reduced ductility $RA_H$ in hydrogen, leading to a larger value for the strength or ductility loss index. HE may also cause faster fatigue crack growth. This is shown in **Figure 2**C for an austenitic steel, which required around half the number of cycles in the presence of hydrogen to reach the same crack length as sample without hydrogen [66]. Strength and ductility loss indices are used throughout this paper to compare HE susceptibility, although we note issues with their suitability as HE metrics when the applied stress intensity factor is greater than the threshold stress intensity factor, which can lead to sub-critical crack growth [67–70].

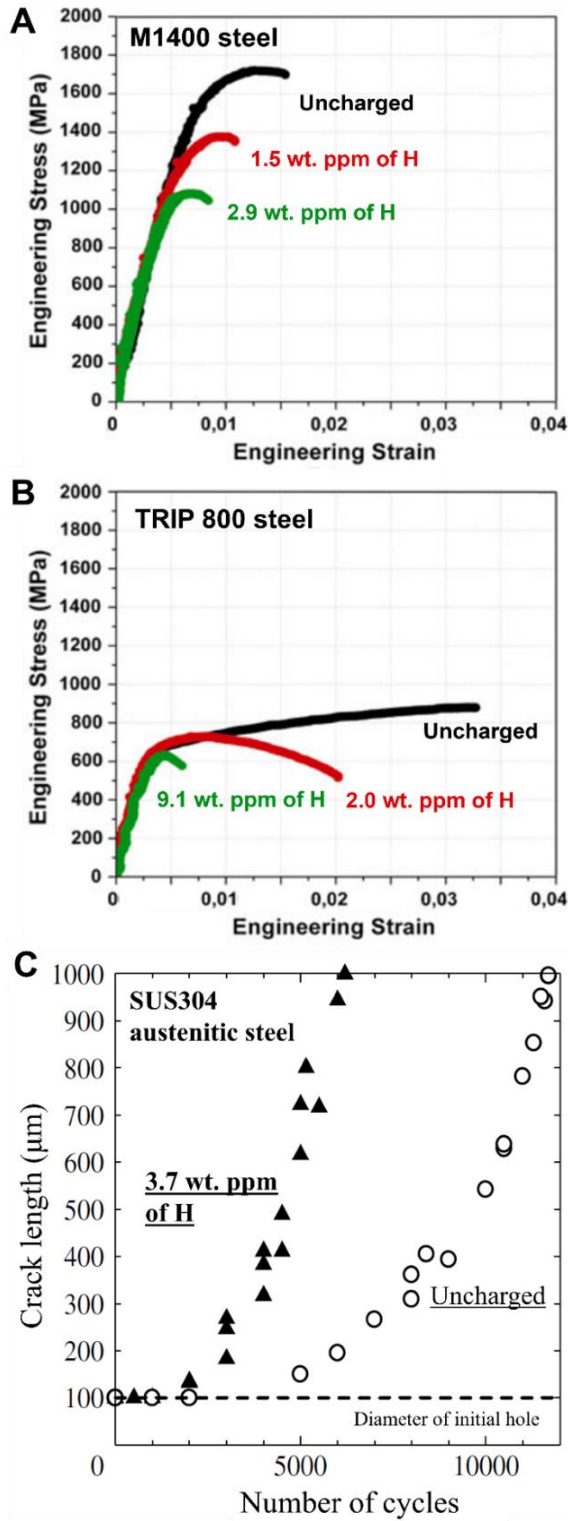

**Figure 2. The effect of HE on strength, ductility, and fatigue resistance.** HE of (A) M1400 martensitic steel and (B) TRIP 800 ferritic/bainitic/austenitic showing reduction of UTS and elongation in a slow strain rate test. The samples were ex-situ charged with hydrogen and the hydrogen content was measured by thermal desorption analysis. (A) and (B) are reproduced from [60]. (C) HE of a SUS304 austenitic steel in a constant-loading fatigue test. The mechanical loading was applied at an amplitude of 280 MPa in both tension and compression and a frequency of 1.2 Hz. (C) is reproduced from [66].

HE can also manifest as hydrogen-induced delayed fracture, which can be characterized by the time-to-fracture in a U-bend test under constant loading, where the time-to-fracture decreases with increasing hydrogen content, stress, and/or strain [71,72]. These tests, which load the specimen below the UTS (hence without immediate failure) indicate that the presence of hydrogen can decrease the threshold stress intensity factor required for crack growth or cause sub-critical cracking for certain stress conditions [31,42,43,73,74]. This type of HE can occur at approximately half of the yield stress with crack velocities up to $10^{-4}$ m/s [73,74]. A threshold time-to-fracture is generally defined (e.g., 300 hours in the case of **Figure 3**A) with a given bending radius (R in the insert figure), and this information informs the time required for manufacturing high-strength steel products in the automotive industry. The U-bend test can be used to create a HE fracture map, as shown in **Figure 3**B, offering a guide for the manufacturing conditions that would avoid HE.

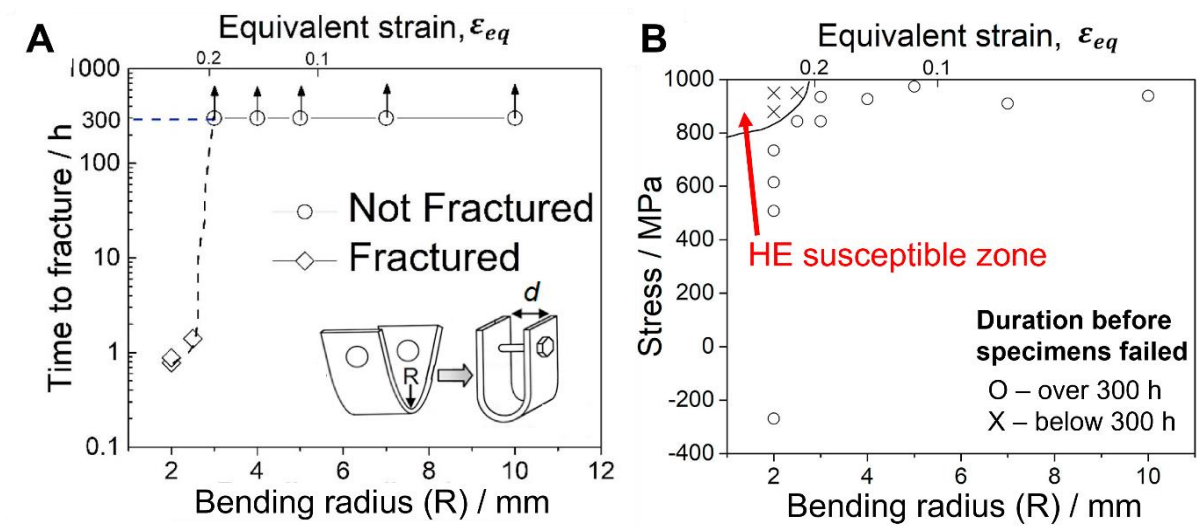

**Figure 3. U-bend test for characterizing H-induced fracture.** (A) Time-to-fracture as a function of strain using a 300-hour limit. (B) Stress-strain HE fracture map with the stress evaluated by either X-ray measurement or finite element calculation. Reproduced from [71]

The effects of hydrogen on alloys are not always detrimental. Exposure to hydrogen occasionally results in an increase in yield strength and ductility, particularly in materials that have higher hydrogen solubility such as face-centered cubic (FCC) iron. An example, shown in **Figure 4**A, is an annealed Fe-24Cr-19Ni-0.02C austenitic FCC steel [75]. **Figure 4**B shows cup and cone fracture surfaces, consistent with ductile dimple rupture. Higher hydrogen levels led to smaller dimples, indicating that hydrogen facilitated ductile fracture. Hydrogen-induced strengthening was attributed to solid-solution strengthening, and the increased ductility was attributed to the facilitation of mechanical twinning by the hydrogen in solution in the austenite [75,76].

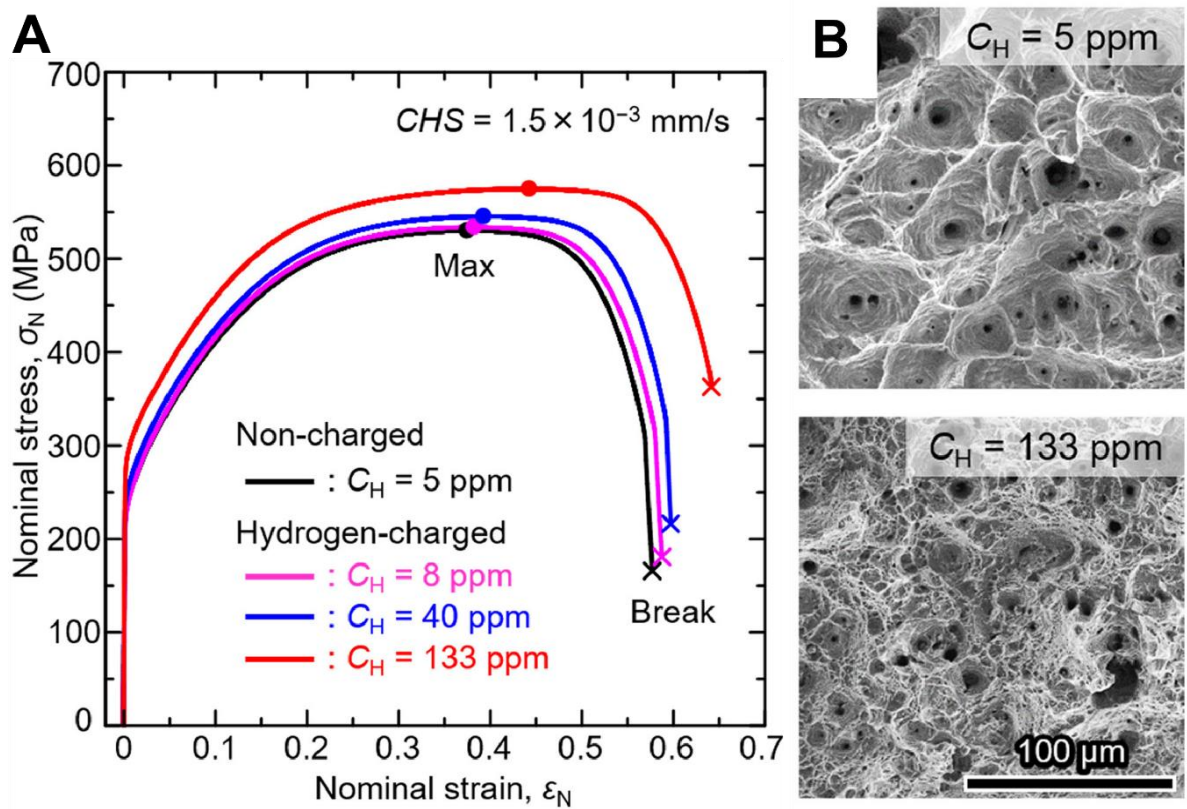

**Figure 4. Hydrogen-induced strengthening and enhanced ductility in an austenitic steel.** (A) Stress-strain curves for an annealed Fe-24Cr-19Ni-0.02C austenitic steel with hydrogen concentrations from 5 wt. ppm to 133 wt. ppm, showing that both strength and ductility increased with increasing hydrogen content. (B) Fractography indicates cup and cone fracture. The specimen with more hydrogen has a smaller dimple size, indicating that hydrogen facilitated the ductile dimple rupture. Reproduced from [75]

Hydrogen has even been proposed as a temporary alloying element to increase the formability of Ti alloys [77,78]. The Ti alloy is held at a relatively high temperature in hydrogen. The absorbed hydrogen induces a phase transformation from α-Ti (hexagonal close-packed, HCP) to more ductile β-Ti (body-center cubic, BCC). The Ti alloy is then mechanically worked at a high temperature and subsequently, the hydrogen is removed by exposure to a low hydrogen fugacity (in a vacuum or an inert gas).

*2.1.3. Fracture mode*

The fracture modes of embrittled alloys are normally characterized by taking scanning electron microscope (SEM) images of the fracture surface from above or in cross-section. Examples are presented in **Figure 5**A (top view) and B (cross-section), respectively, showing hydrogen-induced intergranular fracture along prior-austenite grain boundaries (GBs) in a tempered martensitic steel [31]. Best practice for the examination of fracture surfaces requires observations at a tilted angle and at a high resolution so that critical but subtle HE features can be properly captured and correctly linked with associated mechanisms, as detailed in Section 2.2 [30,31]. For example, **Figure 5**C provides a low-magnification image that may be considered as a manifestation of brittle fracture; however, a high-resolution image from the same specimen indicated the presence of fine dimples, as shown in **Figure 5**D, indicating some plastic deformation during the hydrogen-induced fracture [79]. HE typically causes a macroscopic loss of ductility; nevertheless, there are many microscopic fracture modes including voiding, brittle fracture, and transgranular quasi-cleavage with micro-ridges as shown in **Figure 5**E and F [80]. For ductile fracture, as shown in **Figure 4**, increased density of dimples in HE-affected specimens indicates that hydrogen facilitates micro-void nucleation and growth.

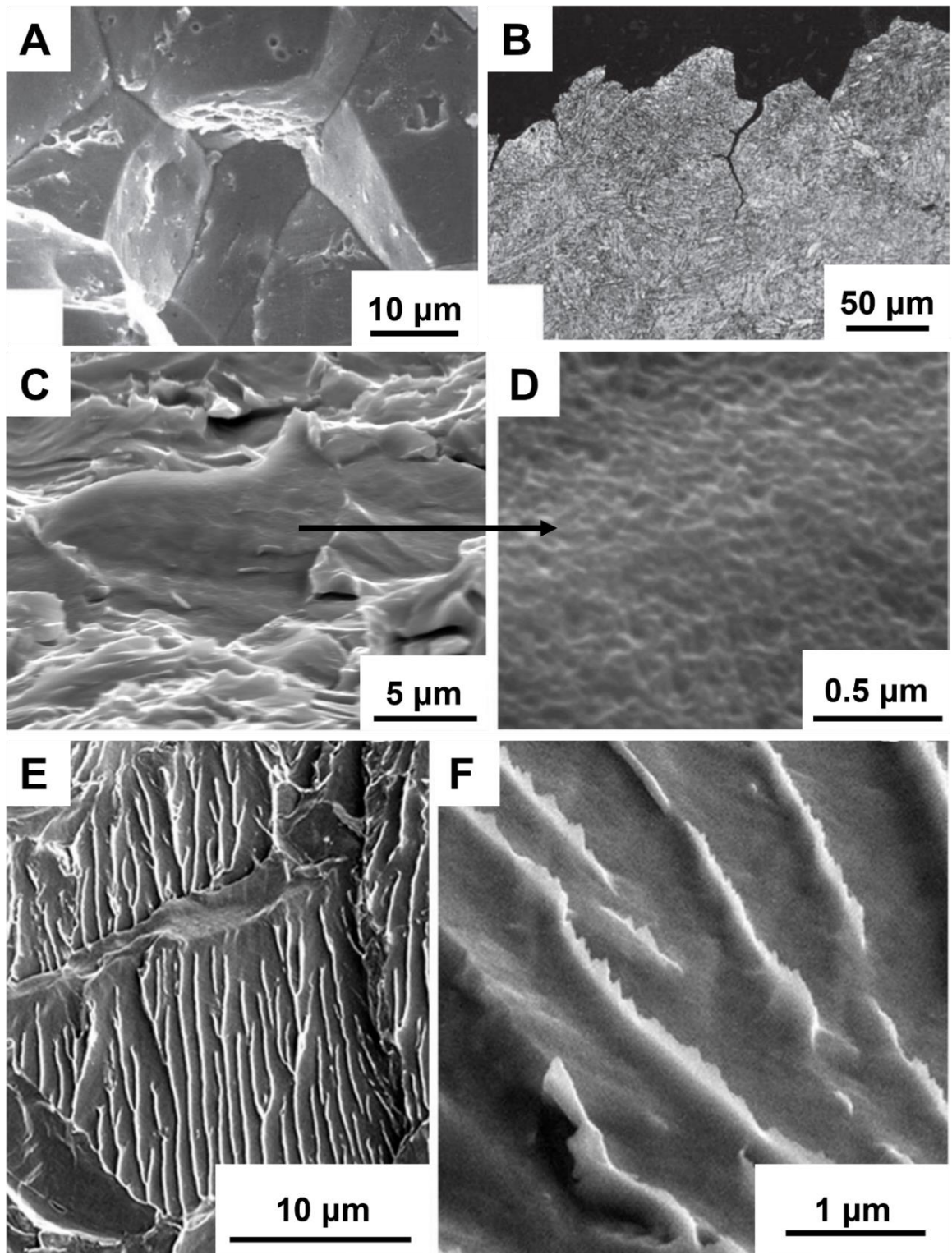

**Figure 5. Fractography of hydrogen-induced macroscopically brittle fractures.** (A) and (B) SEM surface and cross-sectional images of a hydrogen-embrittled tempered martensitic steel, showing mainly intergranular failure due to HE. Source: [31]. (C) and (D) low- and high-magnification SEM images of a hydrogen-embrittled API X60 pipeline steel, showing how tilting and high-resolution imaging reveals dimples on the fracture surface. Source: [66]. (E) and (F) low and high magnification SEM images of a hydrogen-embrittled API X60 pipeline steel specimen, showing the quasi-cleavage fracture surface and its micro-ridges. Source: [80]

*2.1.4. Temperature and strain rate*

HE susceptibility depends on temperature and is often most severe at close to ambient temperature, as shown in **Figure 6** [81]. The influence of temperature is related to the kinetics of hydrogen diffusion and transport toward susceptible areas in the material microstructure [81,82]. At a low temperature, hydrogen does not have sufficient mobility to facilitate the diffusion-controlled HE mechanisms that lead to fracture, although HE can still take place through a mechanism that does not require hydrogen diffusion (such as grain boundary decohesion) [83]. At a high temperature, hydrogen is too mobile to be pinned by dislocations. Similar to temperature, the kinetics of hydrogen movement affects HE susceptibility at different strain rates. **Figure 7**A shows that, in the presence of hydrogen, the UTS and the elongation (ductility) of a martensitic steel are lower (more susceptible to hydrogen) at low strain rates [84]. Fractographic analysis (**Figure 7**B) indicated a higher fraction of the embrittled (intergranular) characteristics at lower strain rates.

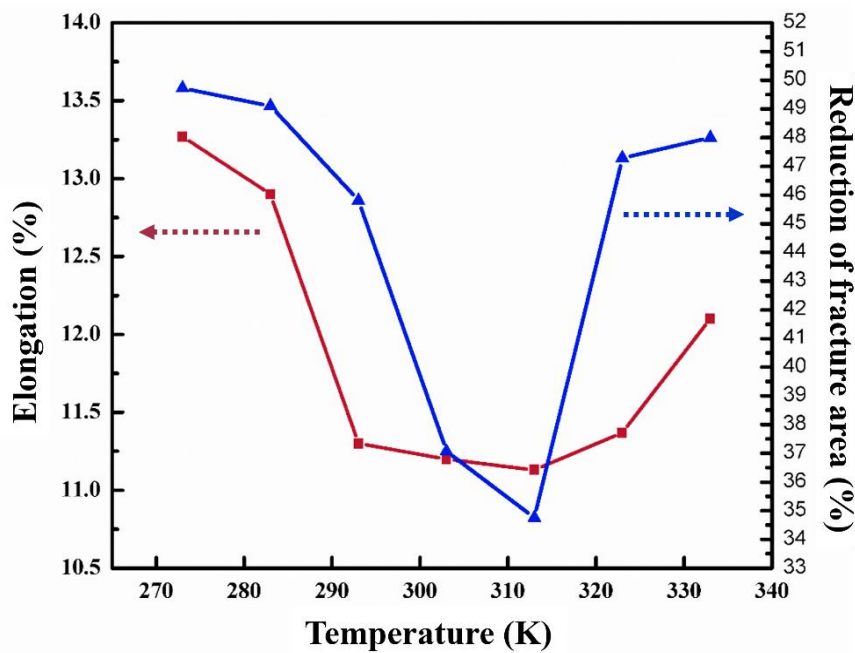

**Figure 6. Temperature window for HE.** The extent of HE of hydrogen-charged ferritic/bainitic X90 pipeline steel specimens in terms of the reductions of elongation (red) and fracture area (blue). The HE reaches a maximum (300-320 K) close to room temperature (293 K). Properties were measured using slow strain rate tests with continuous hydrogen charging. Reproduced from [81]

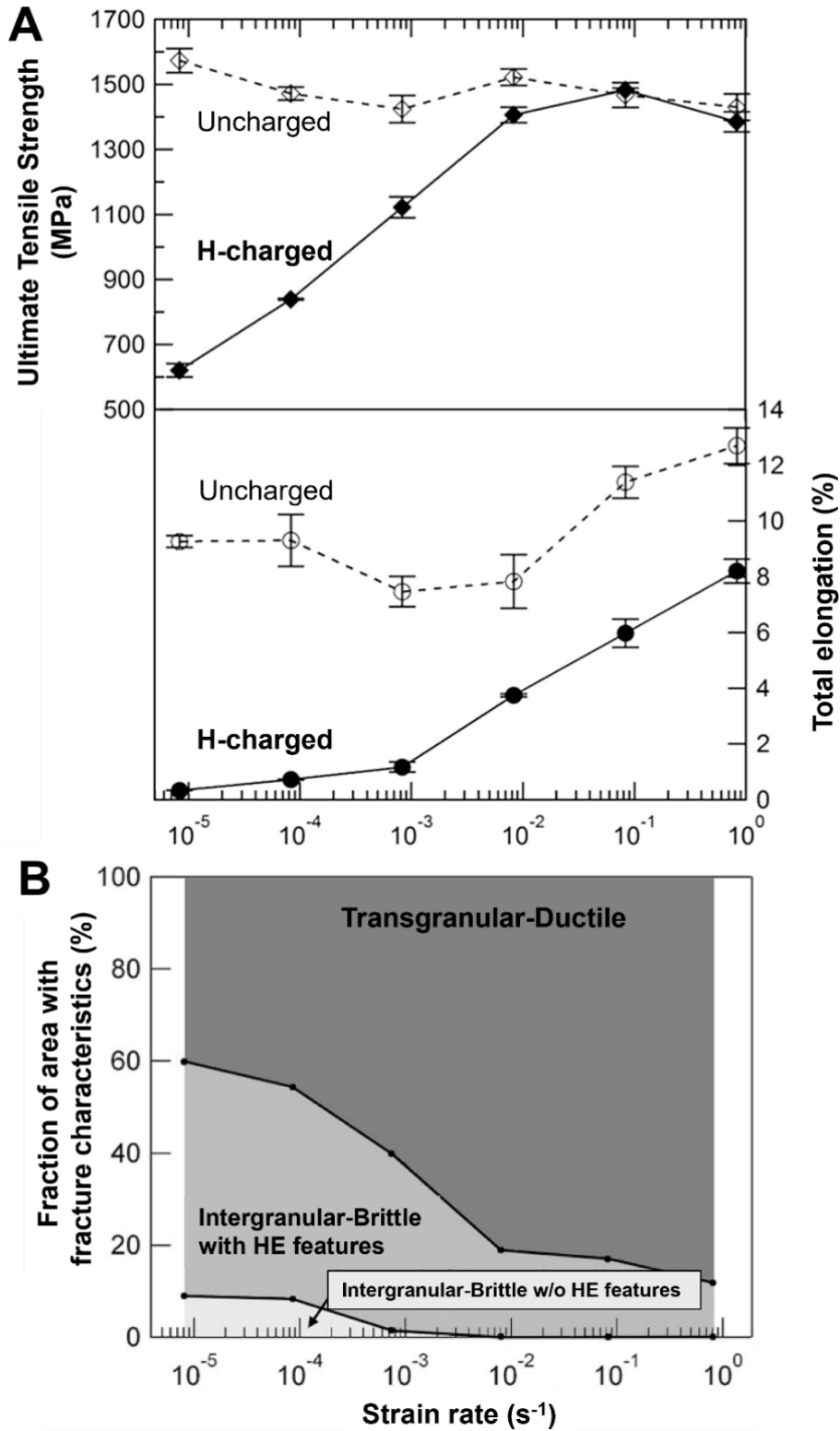

**Figure 7. The effect of strain rate on HE.** (A) The UTS and the elongation of hydrogen-charged and uncharged low-carbon martensitic steel specimens decreased with decreasing strain rate in uniaxial tensile tests, showing HE is greater at lower strain rates. (B) Fractographic analysis of the specimens in (A) showing the surface area fraction of ductile and brittle features, indicating a change from transgranular-ductile to intergranular-brittle fracture with decreasing strain rate. Reproduced from [84]

*2.1.5. Strength*

As a rule-of-thumb, within a class of alloys, the alloys with higher strength tend to be more susceptible to HE [31,35]. As such, extreme care should be taken when considering the use of steel with a strength greater than 1 GPa in a hydrogen-containing environment. This strength-susceptibility relationship can be attributed to the strong interaction of hydrogen with crystal defects such as dislocations and GBs, which largely determine the strength of the material [29–31,36,50,85–88] and higher stresses (and thus hydrogen concentrations) attained in the fracture region [58,89,90]. Correlations of HE susceptibility with dislocation density and grain size, for a wide range of bainitic and martensitic steels with similar strengths, are shown in **Figure 8**A and B, respectively.

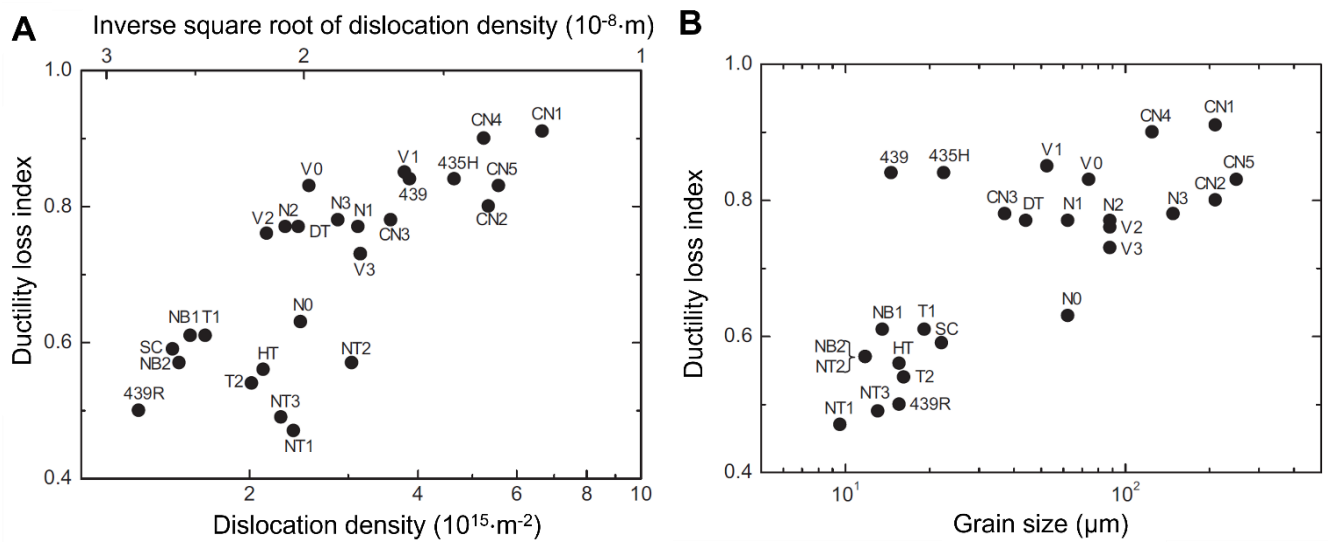

**Figure 8. Dislocation and grain size vs. environmental hydrogen embrittlement susceptibility.** Ductility loss index (HE susceptibility, Equation (2)) as a function of (A) dislocation density and (B) grain size in a range of bainitic and martensitic steels with similar strengths (853-1142 MPa). The ductility differences were measured by comparing the SSRT results in a gaseous hydrogen environment of 45 MPa and in air. Reproduced from [91]

*2.2. Hydrogen embrittlement mechanisms*

Many mechanisms have been proposed to explain HE, based on macro- and micro-scale evidence, but agreement has not been reached on a universally applicable mechanism. In recent years it has become increasingly accepted that several mechanisms may operate simultaneously [29–31,56,92]. The mechanisms described here for hydrogen-induced sub-critical crack growth have also been covered in [29–32,40]. Mechanistic interpretations are used to build predictive models and to provide an explanation for experimental results, particularly for fractography [93–96].

*2.2.1. Hydride formation*

Westlake proposed in 1969 [97] that HE was the result of brittle hydride formation. This embrittlement mechanism is most relevant to alloy systems that have a high tendency to form hydrides such as Zr [41,98,99], Nb [100], Ti [41,101], and Mg [38,102,103]. Hydrogen is an interstitial solute in a metal lattice and can rapidly diffuse and segregate to the zone of high hydrostatic stress at a crack tip (**Figure 9**A). Hydrides form when the hydrogen content exceeds the solubility (**Figure 9**B), and cleavage of the brittle hydride subsequently occurs along the crystallographic direction on which the hydride forms (**Figure 9**C). The crack is arrested when it meets the ductile matrix (**Figure 9**C), and another round of this sequence begins. Crack propagation can be further facilitated by the presence of pre-existing hydrides within the microstructure along the crack path.

In hydride-forming alloys with non-isotropic crystal structure, texture engineering can be used to mitigate HE by promoting the formation of hydrides along crystallographic planes that are less detrimental to the embrittlement of the component. This has been extensively used in the hexagonal Zr alloys, where individual nano-sized hydrides preferentially grow on the basal {0001} plane of the hcp α-phase. Macroscopic hydride plates align approximately on the {10-17} habit plane. In these alloys, tubes and plates are manufactured with a strong "split basal" texture where the c-axis of the hcp crystal is predominantly aligned within 30º of the normal direction of rolling, towards the tangential direction. This encourages hydrides to form parallel to the surface, where they pose the least concern for crack growth. One drawback of this mitigation strategy is that the orientation of hydrides also depends on stress fields, thus upon mechanical loading the hydrides may re-orient across the thickness of the plate or tube, especially after a thermal cycle in which the hydrides are partially or fully dissolved and then re-precipitate upon cooling [98,99].

For non-hydride forming alloys such as Fe or Ni, theoretical models have attempted to relate their transgranular and intergranular failures of HE with the presence of hydrides [104–107]. However, hydride formation at ambient temperature in these alloy systems requires extremely high hydrogen fugacities (e.g., 3.5 GPa for iron [108]), which are unlikely to be relevant in most service conditions.

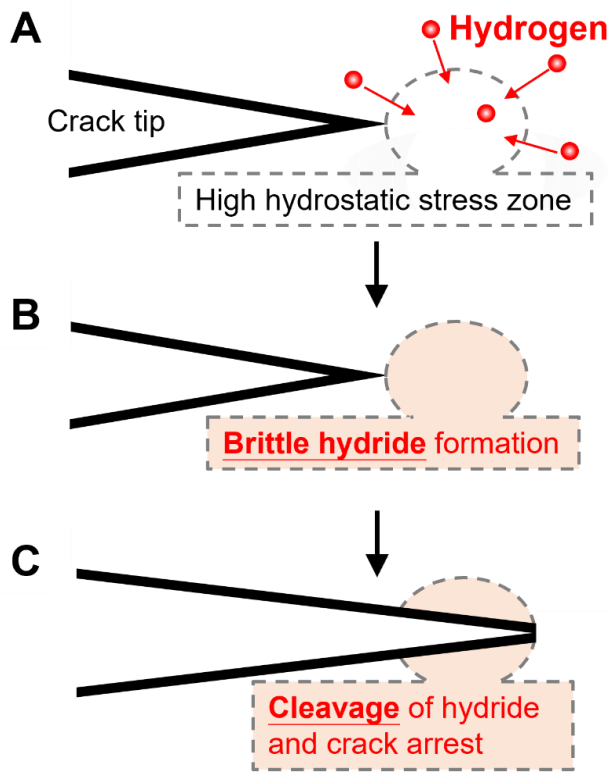

**Figure 9. Schematic of the mechanism of hydride-induced hydrogen cracking for sub-critical crack growth**

*2.2.2. Hydrogen-enhanced decohesion*

In the 1960s and 1970s, Troiano and Oriani proposed the theory of hydrogen-enhanced decohesion (HEDE), which postulates that hydrogen can directly reduce the cohesive strength of the atomic bonds in the metal lattice, leading to brittle fracture, as illustrated in **Figure 10**A [109,110]. While earlier atomistic studies suggested reductions in fracture energy and cohesive strength of up to 90% [111,112], recent studies suggest that any reduction is limited to 20-40% [113–115]. Given that the lattice / grain boundary strength is roughly 10 times the yield stress, a 20-40% reduction in the cohesive strength is insufficient to trigger decohesion in the context of conventional continuum theories [116]. However, theoretical predictions that account for the role of geometrically necessary dislocations and plastic strain gradients lead to much higher stresses and hydrogen levels at the crack tip [117,118], providing a modern mechanistic rationale for HEDE.

HEDE is most widely accepted in the context of inducing intergranular failure (**Figure 10**B) or phase-boundary failure (**Figure 10**C) [117,119,120], particularly for high-strength materials [33,121–125]. Atomistic modeling suggests that the presence of more than one atomic layer of solute segregation can lead to a higher degree of cohesive strength reduction [119]. Experimental evidence of HEDE includes the intergranular failure observed in a hydrogen-charged Ni alloy tested at a cryogenic temperature, where hydrogen-dislocation interactions are effectively suppressed, suggesting that hydrogen-induced GB decohesion is responsible for the HE [83].

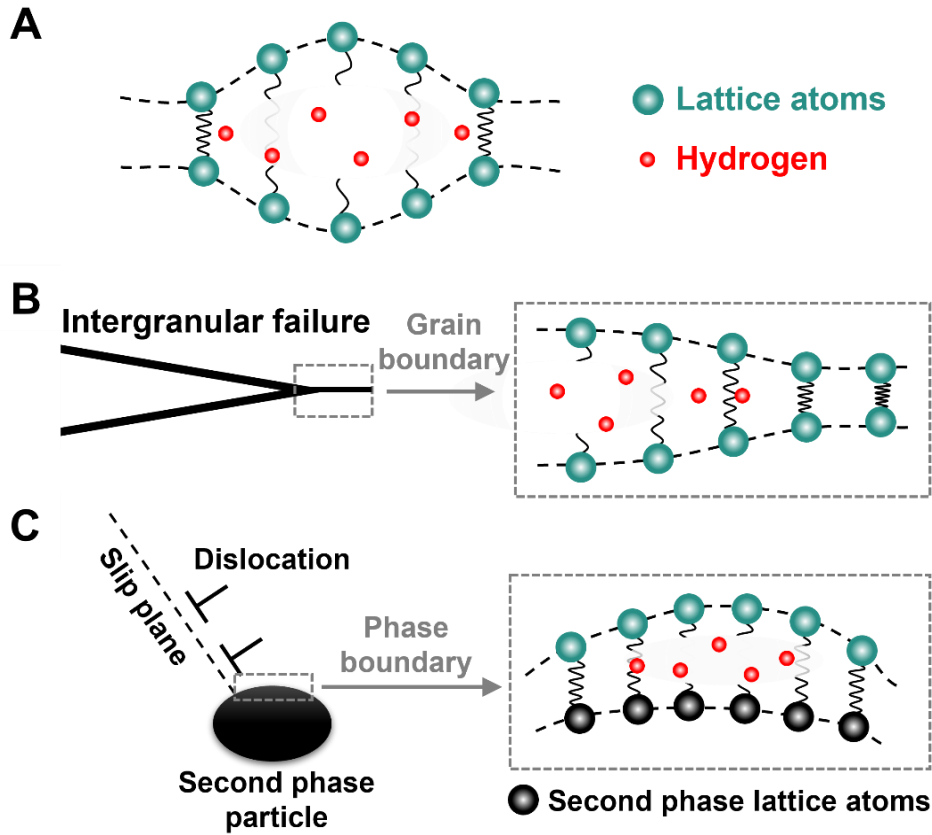

**Figure 10. Schematic of the hydrogen-enhanced decohesion mechanism** (A) in the lattice, (B) at a grain boundary, and (C) at a phase boundary.

*2.2.3. Hydrogen-enhanced local plasticity*

Hydrogen-enhanced ductile fracture with dimpled fracture surfaces (e.g., **Figure 4**B) was described by the hydrogen-enhanced local plasticity (HELP) mechanism by Beachem in 1970s. The theory is that i) interstitial hydrogen atoms concentrate at high tensile hydrostatic stress zones, and ii) hydrogen segregates to lattice defects such as dislocations [126,127], and increases their mobility [128]. This hydrogen segregation is a 'Cottrell atmosphere', a known effect that reduces the strain energy around dislocations [121,129,130]. **Figure 11** illustrates this process. Facilitation of the generation and motion of dislocations promotes the formation of microvoids and their coalescence, allowing sub-critical hydrogen-induced cracks to propagate, resulting in dimpled fracture surfaces.

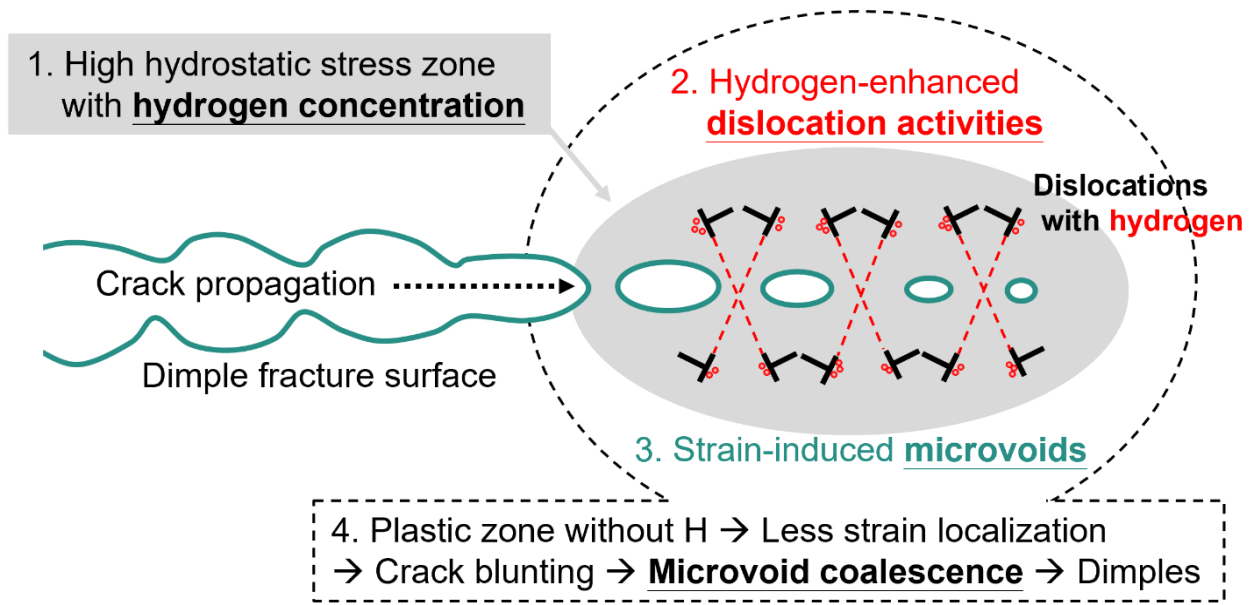

**Figure 11. Schematic of the hydrogen-enhanced local plasticity mechanism**

The central hypothesis of HELP is that hydrogen atmospheres enhance dislocation mobility (Step 2 in **Figure 11**). This was experimentally demonstrated by using an environmental transmission electron microscope (E-TEM) with in-situ mechanical loading on a specimen in a hydrogen atmosphere [128,131,132]. A derivative model of HELP was proposed that associates the crack path with failure along low energy dislocation cell walls that form as a result of the enhanced dislocation mobility in nanoprecipitation-strengthened steels [92]. Noting here again that grain boundary decohesion can still take place in the absence of hydrogen-dislocation interactions in some systems such as nickel alloys [83]. The HELP mechanism is a qualitative description of the effects that lead to embrittlement and does not directly provide a means of evaluating the extent of HE.

2.2.4. *Adsorption-Induced Dislocation Emission*

The adsorption-induced dislocation emission (AIDE) mechanism proposed by Lynch describes HE that arises from hydrogen adsorption at the surface, instead of hydrogen in solution in the bulk [31]. **Figure 12** illustrates the mechanism: HE is caused by the hydrogen adsorption in the metal subsurface [133], which facilitates dislocation activity and boosts dislocation emission from a stressed surface. This dislocation emission facilitates inward strain and leads to the formation of microvoids in the crack-front plastic zone, which coalescence for crack propagation, resulting in a dimpled fracture surface. The AIDE mechanism can explain ductile H-induced crack growth that is induced by an external source of hydrogen, particularly prevalent in specimens with pre-existing surface damage. A similar mechanism, 'hydrogen-assisted micro-fracture', proposed by Atrens and co-workers [134], also attributes HE to the hydrogen effect at specimen surface but stresses that the fracture propagation in the HE of steels (at the UTS) can be as fast as 61-130 m/s, which exceeds the typical velocity of ductile fracture of 46 m/s, suggesting the possible presence of another process that leads to the bulk HE [135].

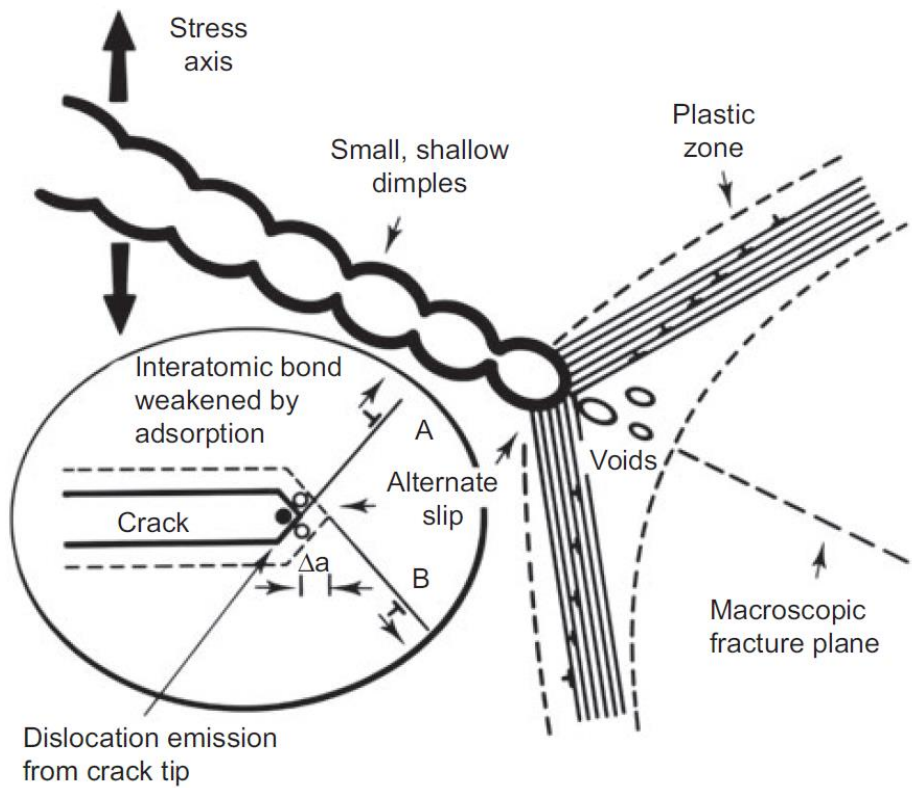

**Figure 12. Schematic of the AIDE mechanism.** Reproduced from [31]

*2.2.5. Hydrogen-enhanced strain-induced vacancies*

Based on the fact that hydrogen facilitates the formation of vacancies [121,129], Nagumo and Takai proposed the mechanism of hydrogen-enhanced strain-induced vacancies (HESIV), which attributes the observed dimpled fracture surface (e.g., **Figure** 4B) to an abundance of hydrogen-stabilized vacancies that are transported by hydrogen-activated dislocations and subsequently combined into nano-voids during deformation [47]. Nagumo combined thermal desorption analyses with delicately controlled straining and annealing to introduce and remove vacancies, respectively, in steel specimens. Hydrogen-vacancy interactions were correlated with HE. Positron annihilation spectroscopy, which can detect both hydrogen and vacancies, was also used to provide evidence of the relationship between hydrogen-vacancy interaction and HE [136–138].

*2.3. Hydrogen uptake and inhibition*

Understanding how hydrogen enters materials, and hydrogen uptake can be inhibited, is essential for developing strategies to reduce HE. This section provides an overview of hydrogen entry, how hydrogen entry can be managed, and introduces methods that relate laboratory test results to HE problems in the field.

*2.3.1. Internal hydrogen embrittlement: hydrogen uptake and liberation*

Hydrogen is a byproduct of many material fabrication processes, and hydrogen uptake can lead to IHE [130,131]. An example occurs in the application of an aluminum-silicon (Al-Si) alloy coating, a common component of the automotive hot-stamping steels [141–143]. Al and Si oxidation in the presence of ambient moisture produces hydrogen, some of which dissolves into the steel substrate [144]:

$$2Al + 3H_2O \rightarrow Al_2O_3 + 6H \qquad (3)$$

$$Si + 2H_2O \rightarrow SiO_2 + 4H \qquad (4)$$

This problem is of particular concern in the automotive industry, where high-strength steel is desirable for weight reduction [144]. Another example is the exposure of hot or molten steel to air humidity during steel production. Hydrogen can be generated and absorbed by the steel when the surfaces encounter ambient moisture at high temperatures [112,115]:

$$3Fe + 4H_2O \rightarrow Fe_3O_4 + 8H \quad (5)$$

$$2Fe + 3H_2O \rightarrow Fe_2O_3 + 6H \quad (6)$$

$$Fe + 4H_2O \rightarrow FeO + 8H \quad (7)$$

Other treatments that can lead to hydrogen uptake by steels are acidic pickling, electroplating for chromium [146], zinc or cadmium [147–151], welding [46,152–154], and in-service corrosion [63]. A famous example of HE following electroplating is the catastrophic failure of tightening bolts during the construction of the San Francisco Bay bridge span in California, USA, which cost 40 million US dollars for rectification [155].

Hydrogen uptake during steel production can be removed by pre-service baking [147,156]. For example, baking at 200 °C for 10 min was found to be sufficient to remove the pre-existing hydrogen in a hot-stamped martensitic steel specimens with an Al-Si coating [156]. This approach requires consideration of the effect of the heat treatment on the microstructure and mechanical properties, which may be problematic for some high-strength alloys [148,157].

*2.3.2. Environmental hydrogen embrittlement: hydrogen surface entry and inhibition*

Hydrogen in the environment absorbed through the metal surface can lead to EHE [139]. The hydrogen supply can be either finite, leading to conditions similar to those of IHE, or infinite (relative to the solubility in the material at the relevant fugacity), as is the case for gas transmission pipelines carrying hydrogen. Hydrogen from $H_2$ gas molecules in the local environment dissociate, adsorb on the surface, and are absorbed into the metal lattice. The process can be formulated as:

$$\tfrac{1}{2}H_2(g) \leftrightarrow H_{ads} \leftrightarrow H_{abs} \leftrightarrow H_{lattice} \quad (8)$$

The absorbed hydrogen content in the material ($c_H$) follows the modified Sieverts' law relating to the solubility constant ($S$), which is specific to the material and its surface condition, the hydrogen pressure or fugacity ($f_{H2}$) in a material, and the amount of pre-existing and trapped hydrogen ($c_T$):

$$c_H = c_T + S\sqrt{f_{H_2}} \qquad (9)$$

The hydrogen fugacity is the pressure that hydrogen would apply if the hydrogen acted as an ideal gas and is essentially equal to the hydrogen pressure below 200 bar. **Figure 13** indicates that the hydrogen concentration in the steels follows the modified Sieverts' Law for hydrogen pressures up to 200 bar [158–160]. Hydrogen traps are filled rapidly with hydrogen as it becomes available (i.e., at hydrogen pressures less than 1 bar), which explains the apparent y-axis intercept in Equation (9) and **Figure 13**. The low hydrogen content in the annealed iron sample indicates that the intrinsic hydrogen solubility in pure iron is quite low, and that the more substantial hydrogen solubility in steels is associated with hydrogen in traps. Sieverts' law, i.e., Equation (9), can be derived using statistical mechanics, as shown in [158], assuming i) the hydrogen in the steel is in equilibrium with the hydrogen in the gas phase, ii) $N_H$ hydrogen atoms occupy $N_S$ sites in the steel lattice having $N_t$ trap sites; and iii) only one hydrogen atom can occupy each of the $N_S$ sites. Note that in many applications (e.g., pipelines, pressure vessels, pumps) the metal is exposed to $H_2$ gas for prolonged periods (relative to temperature and fugacity) and the dissociation/adsorption/absorption reaches a dynamic equilibrium (Reaction (8)). This means that hydrogen atoms recombine at the surface and the dissolved hydrogen also desorbs back out of the surface. This is not the case for alloys exposed to transients of high hydrogen fugacity but is the case when the source of environmental hydrogen is from corrosion of the alloy.

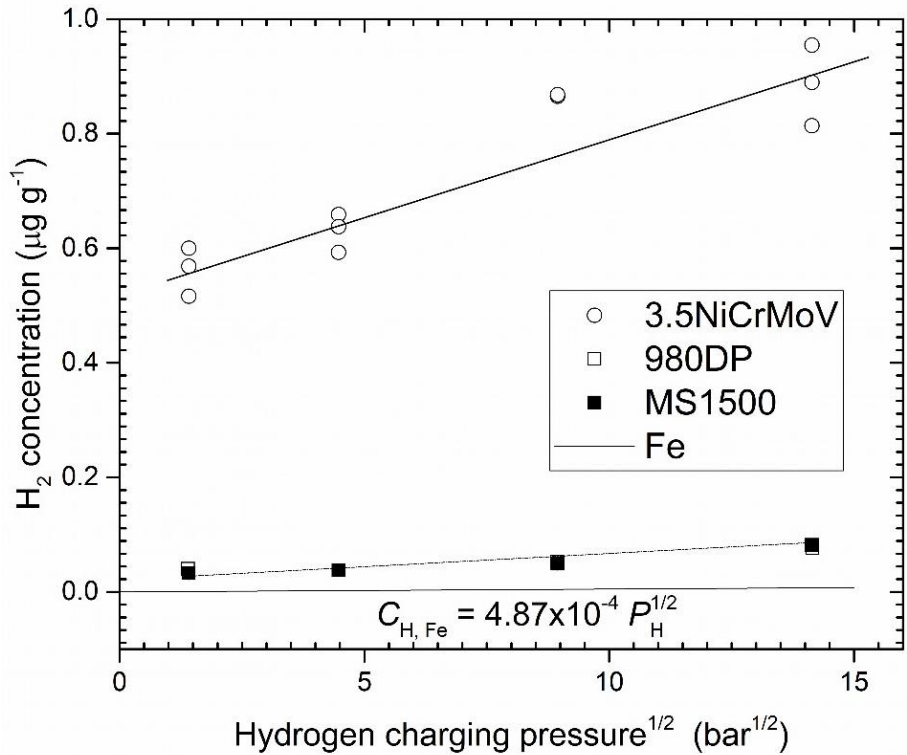

**Figure 13 Equilibrium hydrogen concentration versus charging pressure** in 3.5NiCrMoV steel, 980DP and MS1500 steels for various pressures of gaseous hydrogen from 1 to 200 bar at room temperature, measured using gaseous hydrogen charging [158–160], compared with literature data for annealed pure Fe [161].

In addition to the simplest case of direct contact with gaseous hydrogen, contact with hydrogen-bearing gases such as hydrogen sulfide ($H_2S$) can also lead to HE [17,44,53,59,162]. For an inert gas like methane, the hydrogen solubility in the steel can be determined by Sieverts' law using the hydrogen partial pressure in the gas phase. The hydrogen solubility is much higher (up to an order of magnitude higher) in the presence of an active gas like hydrogen sulfide that inhibits the recombination of hydrogen atoms into hydrogen molecules at the metal surface, thereby increasing the hydrogen fugacity at the metal surface. For this reason, HE due to hydrogen sulfide is a major concern in gas and oil pipes. The usual approach to tackle $H_2S$ embrittlement is to impose a hydrogen gas pressure in service that is one or several orders of magnitude lower than the level that causes HE [163]. An alternative approach is to use thicker pipe walls of low-strength steels that are less susceptible to HE, but this can lead to reduced gas transmission efficiency and results in higher weight and installation costs [163].

In contrast, some gases such as oxygen ($O_2$), carbon monoxide (CO), and sulfur dioxide ($SO_2$) can inhibit EHE. **Figure 14**A shows that the fracture toughness of X42 and X72 steels in gaseous hydrogen was significantly lower than that measured in inert gases like $N_2$ and $CH_4$. However, when the gases were combined, the fracture toughness in hydrogen + CO (indicated by the red arrows) was similar to the fracture toughness in an inert gas, indicating that the CO in the gas mixture inhibited EHE [164]. Similarly, **Figure 14**B shows that $O_2$, CO, and $SO_2$ inhibited hydrogen fatigue crack growth [133]. These inhibitor gases, particularly oxygen, absorb preferentially at active sites on the surface and prevent hydrogen from reaching the steel surface and being adsorbed [165–168]. Based on the compelling laboratory evidence for gas inhibition of EHE, additional research is now required to properly quantify this effect to allow for the use of a gas mixture strategy to address HE problems in gas transmission pipelines.

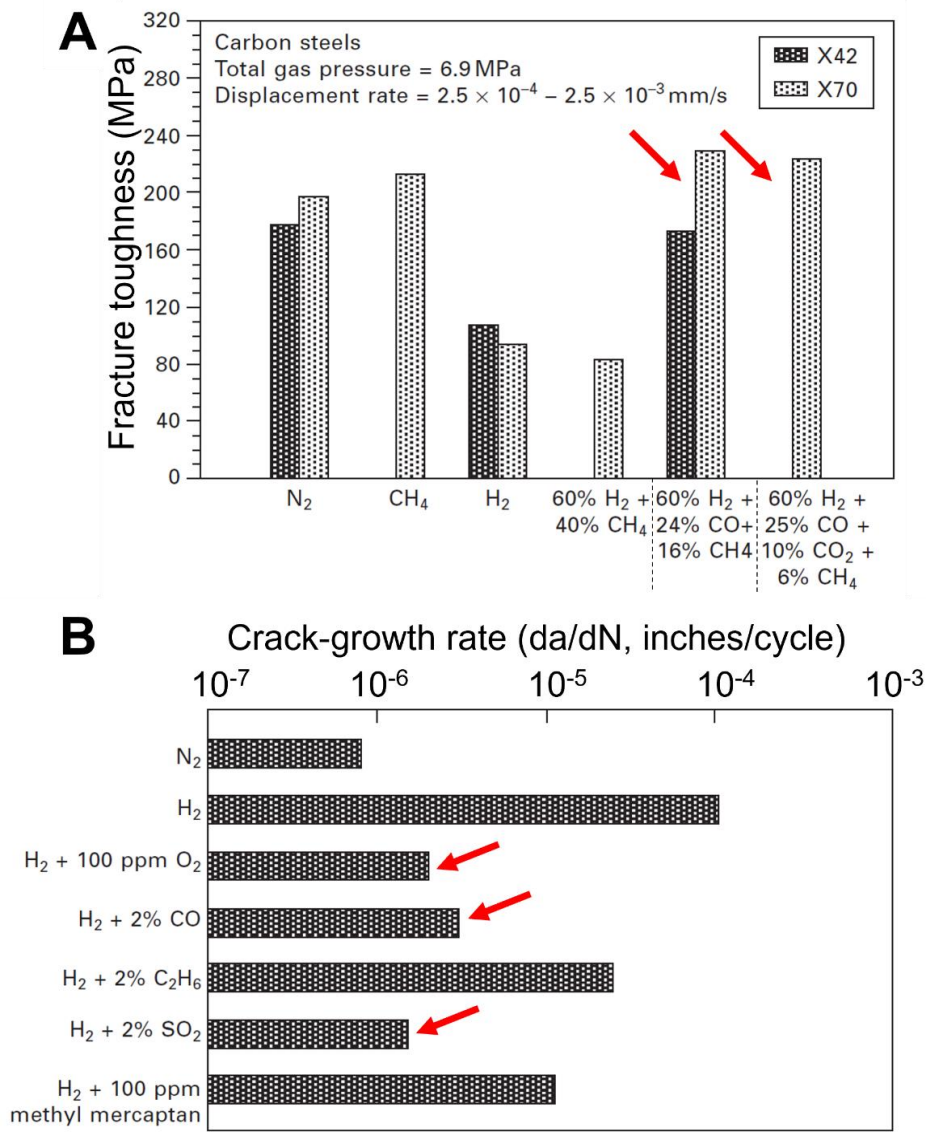

**Figure 14. EHE susceptibilities for pipeline steels exposed to different gas mixtures with hydrogen.** (A) The fracture toughness of X42 and X70 pipeline steels and (B) fatigue crack growth of X42 steel in the gases that contain various contents of hydrogen, showing the inhibition of EHE by $O_2$, CO, $SO_2$. Reproduced from [133].

Surface coatings such as iron oxide and aluminum oxide have also been found to reduce EHE by preventing hydrogen entry [28,169–173]. This approach has the advantage that it can be applied to existing structures. Hydrogen is an effective reductant, so surface oxides tend to be reduced by hydrogen, which may limit the coating life. However, this reduction reaction occurs slowly at room temperature [174]. Metallic coatings and mechanical surface treatments can also be used to increase HE resistance, as demonstrated in [112,144] and [176–178], respectively. However, coatings can crack (particularly in aggressive environments), so surface modification is ineffective for abrasive environments such as bearings [28]. In addition, surface diffusion barriers and gas mixtures only reduce the kinetics of hydrogen uptake and do not alter the thermodynamic factors [158]. Unless hydrogen entry kinetics are sufficiently low, materials will eventually become saturated with hydrogen with continuous hydrogen contact, such as in a gas transmission pipeline that might be used for over 30 years .

### 2.3.3. *Laboratory hydrogen charging*

Laboratory tests use in-situ (concurrent) and ex-situ (prior) hydrogen charging [31,45] to simulate service environments. In-situ charging during a mechanical test (and often also beforehand to achieve a uniform hydrogen content) is optimal for evaluating HE susceptibility. Ex-situ charging is simpler to achieve in practice, but hydrogen loss can affect the quality of results for materials with high diffusivity or where long-duration tests (e.g., fatigue) are required. Erroneously high HE resistance can result from hydrogen depletion at the surface, where hydrogen cracking typically initiates. During hydrogen charging, surface oxides present on the metal surface may prevent hydrogen uptake, which can be mitigated by an activation procedure.

Gaseous charging generally requires a high pressure and/or high temperature to enable hydrogen uptake [179]. This approach is thus demanding in terms of laboratory safety and equipment. In contrast, cathodic charging requires only an electrolytic cell, which allows the specimen (the cathode) to be charged in an acidic, neutral, or alkaline solution [149]. Cathodic charging allows for easy generation of high fugacity hydrogen (at the level of GPa) within the specimen with an overpotential of less than 1 volt [158,180,181]. Much higher hydrogen fugacities are possible if a hydrogen recombination poison is also added to the cathodic charging solution. Depending on the charging parameters, the resulting microstructure can display a strong hydrogen concentration gradient within the samples with a hydrogen-rich (or even hydride) surface and a hydrogen-poor core, thus cathodic charging is often followed by a low-temperature heat treatment to redistribute the hydrogen within the bulk specimen. The simplicity and effectiveness of cathodic charging have led to its widespread use for HE testing [151], despite the fact that quantifying hydrogen uptake is not as straightforward as it is for gaseous charging. Hydrogen can also be introduced by charging from a plasma [183,184], but this approach is less common and it remains uncertain whether the low pressure required allows for the introduction of sufficient hydrogen to trigger HE [185,186].

To better understand the effective cathodic charging fugacity, Atrens and co-workers developed a method to relate the hydrogen fugacity to the equivalent hydrogen pressure during gaseous charging [158–160,180]. This approach is based on the assumptions that i) the hydrogen inside the metal has no memory of whether it originates from gaseous charging or cathodic charging and ii) the fugacity during cathodic charging is equivalent to the fugacity during gaseous charging for the same hydrogen concentration. The hydrogen fugacity ($f_{H_2}$) for pure iron is related to the applied electric potential ($E_c$) by:

$$f_{H_2} = \alpha \, exp\left[\frac{-(E_c - E_H^0)F}{\beta RT}\right] \qquad (10)$$

where F, R, and T are the Faraday constant, gas constant, and absolute temperature, respectively. $E_H^0$ is the equilibrium potential of the hydrogen evolution reaction at the sample surface for hydrogen charging at 1 atm of fugacity, hence $(E_c - E_H^0)$ is the overpotential that controls the hydrogen fugacity. The constants α and β relate to the hydrogen evolution reaction (i.e., the acidity of the charging solution) at the sample surface. These constants can be determined for pure iron, which does not have hydrogen traps, based on literature value of the Sieverts' constant $S$ in Equation (9) (**Figure 13**). For a steel, the hydrogen solubility data (e.g., **Figure 13** in which hydrogen concentration is measured for each cathodic charging condition) can be used to experimentally establish the relationship between hydrogen content and hydrogen fugacity [180]. These studies have been further improved by electro-chemo-mechanical models that resolve all the reaction rates involved in the hydrogen evolution reaction [187].

This quantification provides the hydrogen fugacity (and equilibrium hydrogen content) at the specimen surface. Time is required to establish a uniform hydrogen concentration throughout the specimen. The penetration of hydrogen can be predicted using hydrogen diffusion equations. Often measurements of the overall hydrogen content in the specimen are used to assess the rate of penetration. In addition to the charging parameters, the hydrogen content in the specimen can be influenced by the surface condition, particularly surface oxides, which can affect the hydrogen uptake [158]. Another important experimental consideration is time-to-test after hydrogen charging, which affects the extent of hydrogen desorption [188]. Even quite short times can alter the measured degree of HE in some materials. Where possible, in-situ charging during mechanical testing is ideal. When reporting HE results using hydrogen charging, it is necessary to give a clear and complete description of the experimental conditions to enable comparison with the literature.

## 3. Hydrogen trapping

In addition to *extrinsic* approaches to mitigate HE hydrogen by degassing or surface coating, a popular *intrinsic* approach is to introduce additional hydrogen traps into the alloy microstructure. Pressouyre [139] proposed the concept of introducing microstructural hydrogen traps to mitigate HE in the 1980s. Certain microstructural features act as hydrogen traps, which may reduce the amount of hydrogen that is available to reach the crack tip. This concept assumes that the trapped hydrogen is passive and that it is the lattice hydrogen or diffusible hydrogen that leads to HE [27,28,31,32]. However, as shown in **Figure 13**, the presence of hydrogen traps increases the solubility of hydrogen in all alloys, including commercial steels, and these sites can therefore provide a future source of hydrogen available for HE.

An optimal microstructure to resist hydrogen embrittlement is one in which the traps do not contribute to embrittlement themselves, nor do they provide an easy source of hydrogen that can contribute to embrittlement effects. Here we review the fundamental understanding of hydrogen trapping in metals and alloys to establish a scientific ground to underpin the design of a HE-resistant microstructure, discussed in the final section.

*3.1. Principle*

A hydrogen atom in solution in an alloy can either be a free-moving solute in the lattice or it can reside in a trap. **Figure 15** presents the energy states for a hydrogen atom: i) the activation energy needed for classical migration of hydrogen in a lattice site ($E_d$), ii) the activation energy needed for migration close to a trapping site ($E_s$), iii) the de-trapping energy needed for a trapped hydrogen atom to diffuse in the lattice ($E_a$), and iv) the binding energy of the hydrogen atom to a trapping site ($E_b$). These energies can be understood in terms of a simple reaction model where the rate of detrapping $\frac{dx}{dt}$ is given by [189]:

$$\frac{dx}{dt} = A(1-x)exp\left(\frac{-E}{RT}\right) \qquad (11)$$

where $x$ is the fraction of free (detrapped) hydrogen, and R and T are the gas constant and absolute temperature, respectively.

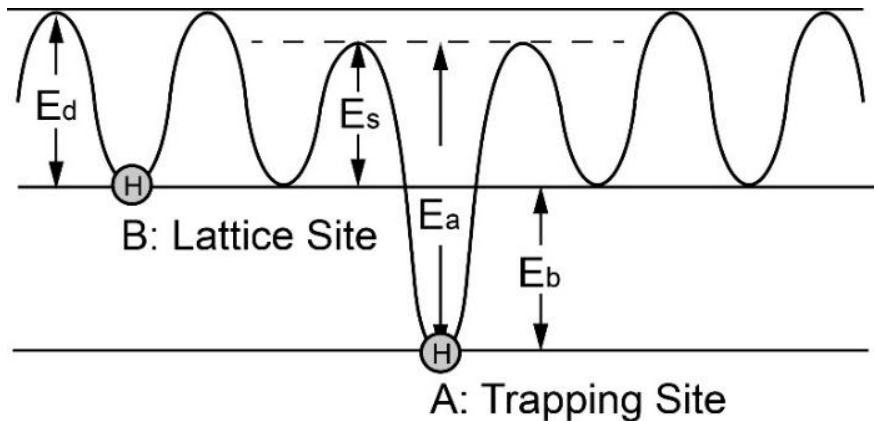

**Figure 15. Schematic energy profile of hydrogen in the lattice and in a trap.** Classical lattice diffusion of hydrogen requires $E_d$ to hop between interstitial sites. This may vary near trapping sites to $E_s$. Detrapping of hydrogen requires an energy of $E_a = E_b + E_s$.

For BCC structures at room temperature, interstitial hydrogen atoms preferentially occupy tetrahedral sites rather than octahedral sites [190–192]. In FCC alloys at room temperature, hydrogen atoms tend to reside in octahedral sites, which are larger and have lower multiplicity for hydrogen migration [190]. This leads to a lower diffusivity and a larger $E_d$ in FCC crystals (e.g., 44 kJ/mol for FCC iron [193]) compared to other crystal lattices (e.g., 4.5-5.5 kJ/mol for BCC iron [194]).

Negative trapping energies reflect that the hydrogen atom in the potential well of the trap is more stable and remains in the trap longer than it does in a lattice site. Both $E_a$ and $E_b$ in **Figure 15** have been defined as the 'hydrogen trapping energy' in the literature, depending on the measurement techniques. The difference ($E_s$, differs from $E_d$) is due to local atomic-structural heterogeneity near the microstructural trap with respect to an ideal lattice, such as the strain field around a dislocation [195]. $E_s$ may also be affected by multiple trapping sites in a single microstructural feature, e.g., the interface and the precipitate structure in second-phase precipitates [196–198].

The value of $E_a$ is often used to indicate the reversibility of a hydrogen trap. At room temperature, a trap with $-E_a$ larger than 50 kJ/mol is considered as 'irreversible' since there is a negligible probability of escape of a trapped hydrogen atom [28,139]. Irreversible traps act as sinks until they are saturated, whereas reversible traps can act either as sinks or sources of hydrogen. For example, if a dislocation (a mobile trap) passes a reversible trap that has a trapping strength lower than that of the dislocation, then the hydrogen in the reversible trap can be transferred to the dislocation and be carried along with the dislocation [139]. Hydrogen trapping/de-trapping is a thermally activated process, so a higher trap strength is needed to confine hydrogen at a higher temperature. This is an important consideration in certain application environments, such as in the components of a nuclear reactor [199].

Considering that hydrogen is highly mobile, equilibrium between the hydrogen in the lattice and in trapping sites can be reached rapidly in most circumstances [42,43]. The relationship between mobile and trapped hydrogen concentrations can be plotted against trap strength, as shown in **Figure 16** [200]. For a $10^{-2}$ wt. ppm hydrogen in the BCC iron lattice (at the bottom of **Figure 16**), the strong traps with 50 kJ/mol trapping energy have a > 90% occupancy ($\theta_T > 0.9$); whereas weak traps with 10 kJ/mol are less than 10% full ($\theta_T = 0.1$), so that most of the hydrogen is in the lattice.

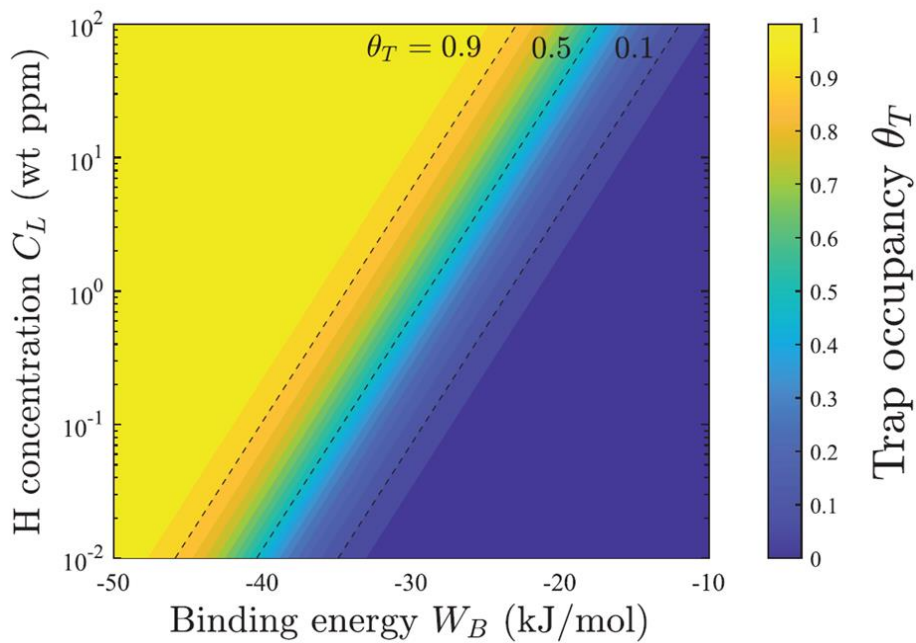

**Figure 16. Relationship between hydrogen content and trap occupancy as a function of trapping/binding energy at absolute temperature.** As the lattice hydrogen concentration ($C_L$) increases, traps with low binding energies ($W_B$) are filled and the trap occupancy ($\theta_T$) increases. Reproduced from [200]

Finally, it is important to note that **Figure 15** depicts the classical hopping mechanism of atomic migration. Hydrogen is sufficiently light that quantum tunneling, and nuclear quantum effects can play a significant role in migration [195,201–204]. Recent computational work showed that quantum tunneling can result in a non-linear Arrhenius diffusivity of hydrogen in BCC iron up to the temperature range of 400–500 K, reveal that quantum effects play a crucial role in the process of H migration even at ambient temperatures [205,206]. Nuclear quantum effects were also demonstrated to influence hydrogen solubility, mobility, and trapping in the presence of strain [206,207]. It has been shown that nuclear quantum effects can increase the rate of trapping and decrease the rate of escape when hydrogen atoms interact with vacancies in BCC iron at low temperatures [182]. Similar findings were obtained in diamond [209–211], and it is likely that the consideration of quantum effects are essential for the behavior of hydrogen in other material systems. Quantum effects should be considered in the study of hydrogen behavior, especially in cases where the effects are most significant, such as in low temperature hydrogen trapping, high-pressure phase stability, mixed hydrogen isotopic composition, and hydrogen migration [212,213].

Overall, hydrogen behaves similarly to other interstitial elements in a metallic matrix, i.e., it segregates to any location where its location causes the energy to be lower [214,215], or where atoms are dragged by vacancies, dislocations, or other mobile defects. In fact, 'hydrogen trapping' highlights the uniquely high mobility of hydrogen compared to other interstitial solute atoms with larger atomic volumes and masses. **Figure 17** illustrates the wide array of microstructural features of engineering alloys that can act as hydrogen traps, for several length scales. Hydrogen trapping can take place at many feature types, including dislocations, vacancies, second phase interfaces, and gain boundaries [216].

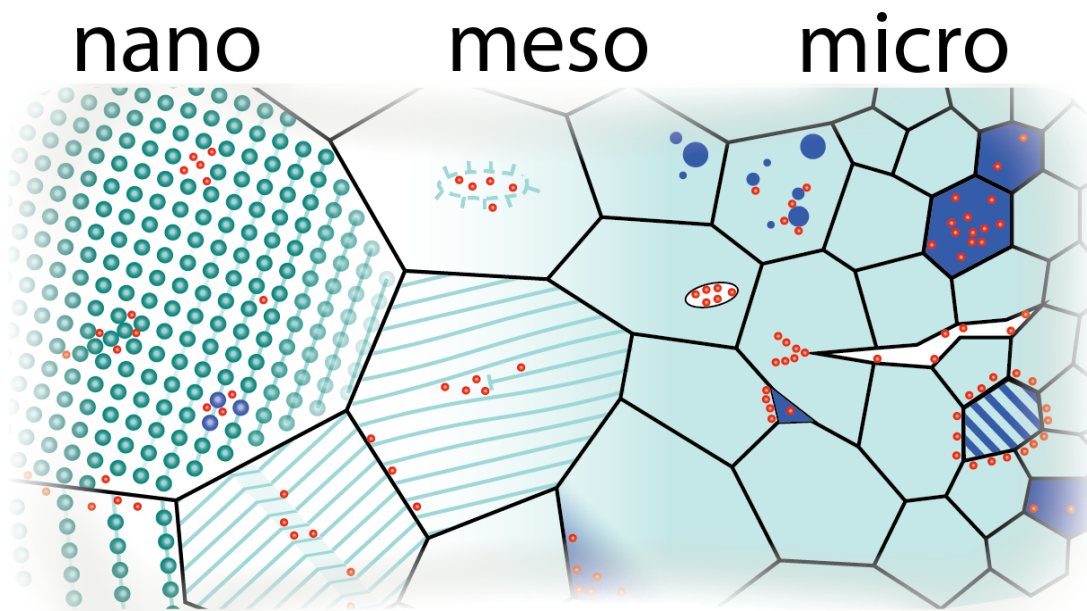

**Figure 17. Illustration of various microstructural hydrogen traps in an alloy.** Hydrogen

can be trapped at interstitial lattice sites, grain boundaries, vacancies, alloying solutes, stacking faults, twins, dislocations and their cell walls, strain field, voids, second phases and their boundaries, and the free surfaces of microcracks.

*3.2. Techniques for studying hydrogen trapping*

This section introduces the computational and experimental techniques that are used for understanding hydrogen trapping. An overview of the techniques is followed by typical results for various common hydrogen traps. This establishes the foundation for the subsequent discussion about the roles of hydrogen traps in developing HE-resistant materials.

*3.2.1. Atomic-scale modeling*

The performance of atomic-scale simulations of hydrogen in metals requires consideration that hydrogen often displays complex chemical binding, especially with transition metals. Additionally, hydrogen is amphoteric and has charge states of –1 (hydride ion), 0 (atomic hydrogen) and +1 (proton). Thus, atomic-scale simulations of hydrogen require a level of theory capable of capturing these nuanced chemical interactions. For this reason, first-principle simulations using density functional theory (DFT) have been the main approach for understanding hydrogen tapping, as they rely on a quantum mechanical description of the electronic system. DFT simulations, either static or dynamic, provide both thermodynamic and kinetic information by determining the ground state energy of hydrogen in crystal lattices and at traps, and the activation energy for migration, dissociation, adsorption, and absorption. The example in **Figure 18**A describes the energy for the absorption of a hydrogen atom into the tetrahedral interstitial site of BCC iron from the (100) surface [217]. The other example in **Figure 18**B compares trapping/solution energies of single and multiple hydrogen atoms in various positions in a ferrite matrix containing vanadium carbide ($V_4C_3$) [218], indicating that hydrogen is preferentially trapped in the carbon vacancy of $V_4C_3$ rather than the interstitial sites of both ferrite and $V_4C_3$. This result also shows that it is possible for a vacancy to accommodate up to two hydrogen atoms. This behavior is observed in many other materials.

In some cases a single vacancy can accommodate up to 12 hydrogen atoms [219–224]. DFT is also used for studying the interaction of hydrogen atoms with alloying solute atoms [192,225,226], grain boundaries [227–229], and second phase precipitates [192,218,230–237].

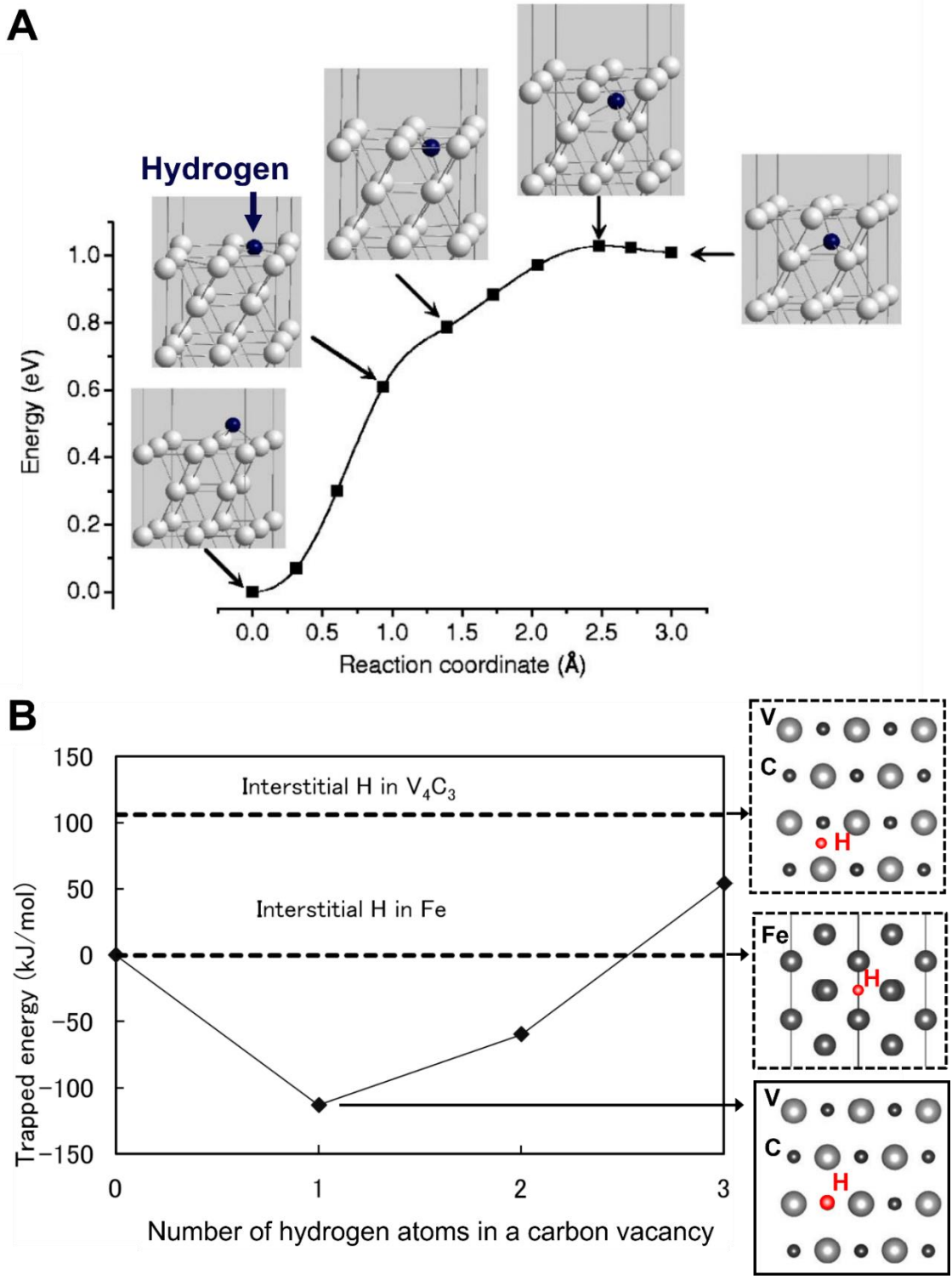

**Figure 18. DFT calculation of hydrogen solute energy.** (A) At the ferrite surface [168]. (B) At an interstitial site or a carbon vacancy of vanadium carbide ($V_4C_3$) [218].

Recent DFT studies have shown that the ground state energy of hydrogen in an interstitial site is strongly dependent on the composition of its immediate surroundings and is relatively insensitive to crystal features and defects beyond the first shell of nearest neighbors [192,226,235,238–241]. This insight is being used to develop a high throughput simulation method based on DFT simulations of the local environment, thus avoiding the computational burden of modeling a large simulation cell representative of the true alloy [192].

The significant computational cost of DFT limits the scale to a few hundred atoms. This size of model does not allow a full description of many of the microstructural features involved in HE mechanisms and hydrogen trapping. Even a simple dislocation requires over 1,000 atoms for a full description of the core and local structure. Longer length scales requires use of classical molecular dynamics (MD) or tight-binding [242], which is an approximation of DFT with a tunable degree of assumptions and thus a sliding scale between DFT and classical MD. However, a recent study of Zr-H potentials showed that reliable Zr-H tight-binding simulations require a similar level of computational cost to DFT [243]. Also, there are few tight-binding codes available, and these are typically not as evolved or approachable as DFT codes. On the other hand, classical MD lacks the chemical accuracy and transferability of DFT because it relies on a pre-defined description of inter-atomic forces. Some of the more complex and advanced interatomic potentials can describe multiple charge states of hydrogen [244,245], but these come at greater computational cost and have shown limited robustness and transferability. Recent developments of machine-learning potentials for MD also show good promise to facilitate HE research [246–248].

In contrast, the hybrid quantum mechanics-molecular mechanics (QM-MM) method is a promising development for larger-scale simulations of hydrogen in metals. This approach divides a simulation cell into two or three concentric regions. **Figure 19** shows an example in which an inner region centered around the hydrogen atoms uses DFT to describe the chemical interactions with quantum mechanical accuracy, while the outer region uses classical potentials to describe the long-range structure of the microstructural feature of interest. An intermediate "buffer" region where both DFT and classical potentials are used is often also added to provide convergence and eliminate artifacts of discontinuity between the two regions. QM-MM has proven successful for molecular and biological systems [249], but it has only recently started to be used for solid-state metallic systems, for example, in the investigation of the interaction between hydrogen and dislocation in BCC Fe [250]. More applications of this new method with the continuing increase of computational power can be expected in near future.

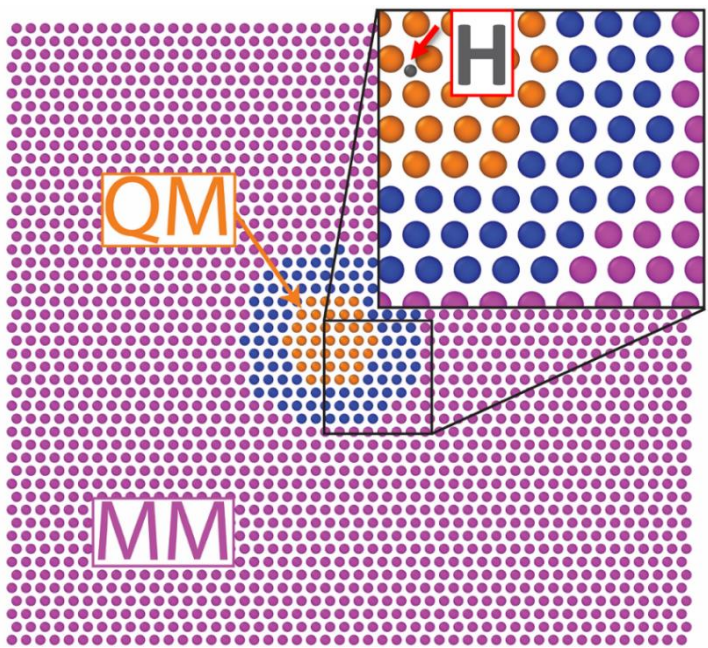

**Figure 19. 2D slice of a MM/QM model to simulate a hydrogen atom.** The small grey sphere in the inset is the hydrogen atom in a screw dislocation in an FCC lattice, where the color indicates the level of theory used to compute forces on those atoms: orange indicates quantum mechanics (typically DFT), purple indicates classical molecular mechanics potentials (e.g., embedded atom method, EAM), and blue indicates a buffer region where both approaches are used to avoid discontinuities.

The methods above can be paired with other techniques when nuclear quantum effects are important. An example is path integral molecular dynamics (PIMD) [251,252], which has made progress in the last decade to make it approachable even at the DFT level of theory [212,253–258]. In PIMD, instead of following one trajectory of motion dictated by Hook's and Newtown's laws, the simulation follows several imaginary trajectories and integrates between them to yield a probability of the hydrogen position and vector. These additional aspects indicate changes to the properties of hydrogen. **Figure 20** shows that the predicted diffusivity of hydrogen in BCC iron can be underestimated by several orders of magnitudes if nuclear quantum effects are not included [207], and this is supported by literature findings [205,206,208]. The same study found that approximately 30% of the hydrogen binding to a vacancy in BCC iron is due to finite temperature effects, and about half of that arises from nuclear quantum fluctuations [207]. In addition to PIMD, the Many-Interactive-World theory of Hall, Deckert, and Wiseman [259], recently implemented by Sturniolo [260], has received increasing attention in hydrogen modeling. This approach complements PIMD and describes the ground state rather than a dynamic state at a finite temperature. These approaches provide a path to include quantum effects in understanding hydrogen trapping.

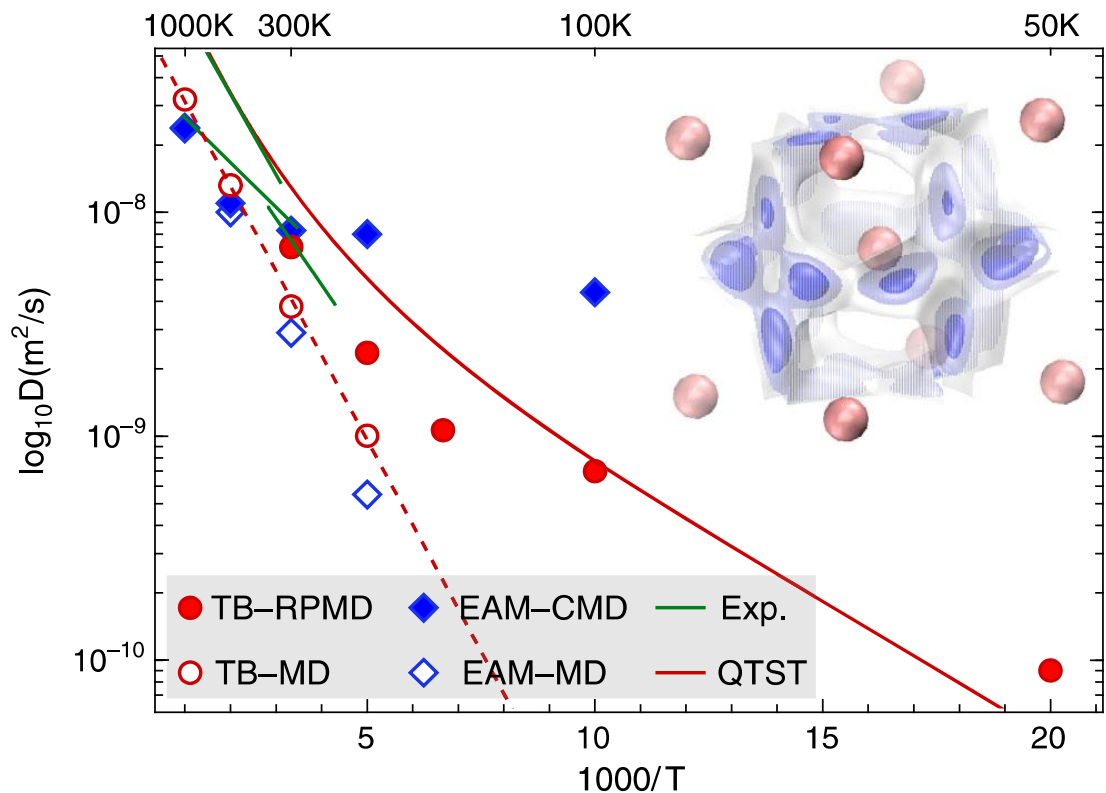

**Figure 20. Effect of including nuclear quantum effects (including tunneling) on the calculated diffusivity of hydrogen in BCC-Fe.** Hollow points only consider classical mechanisms (classical molecular dynamics). Filled points include nuclear quantum effects through path integral methods. Red points use tight-binding to describe inter-atomic interaction from [207], blue points use classical potentials from [205]. The green line is experimental data from [161,261] and the red line is quantum transition state theory (which also accounts for quantum tunneling) from [211]. Inset is the density distribution of hydrogen (blue iso-surfaces) in the BCC-Fe lattice (red sphered). Reproduced from [207]

*3.2.2. Thermal desorption analysis*

Experimentally, thermal desorption analysis (TDA) and hydrogen permeation tests are both used to measure hydrogen trapping. A TDA set-up is illustrated in **Figure 21**A. A hydrogen-charged bulk specimen is placed in a programmable furnace that connects to a gas detector (either a gas chromatograph or a quadrupole mass spectrometer), a vacuum pump to evacuate the desorbed gas, and, in some cases, an inert purging gas (typically nitrogen or helium) to minimize artifacts from residual ambient gas in the chamber [188] and to avoid gas-surface reactions. Desorbed hydrogen is measured as the specimen is heated (**Figure 21**B). Hydrogen desorption peaks at different temperatures are attributed to the desorption of hydrogen from various microstructural hydrogen traps. This is supported by knowledge of the specimen microstructure obtained by using advanced characterization techniques such as transmission electron microscopy (TEM) [262,263]. Hydrogen de-trapping is a thermally activated process, so the de-trapping energy of a peak associated with a specific hydrogen trap can be evaluated by repeating desorption experiments with different heating rates on the same material (but not the very same specimen since the heating can change the surface and microstructure of the specimen) [61]. TDA has also been used in isothermal mode to allow the measurement of hydrogen diffusion at various temperatures [264,265].

Additional insight about the link between specific microstructural features and hydrogen desorption peaks is possible by comparing the TDA of specimens that have different heat treatments or different compositions [180]. However, the assignment of peaks can be speculative [188] and it is common for a large peak to have contributions from multiple overlapping constituent peaks, (e.g., **Figure 21**B), which then requires a deconvolution process to link the constituent peaks with individual microstructural features [182]. Contemporary hydrogen trapping research usually combines TDA data about the trapped hydrogen content and trapping energies with data obtained using complementary analytical techniques [73,162,182,183].

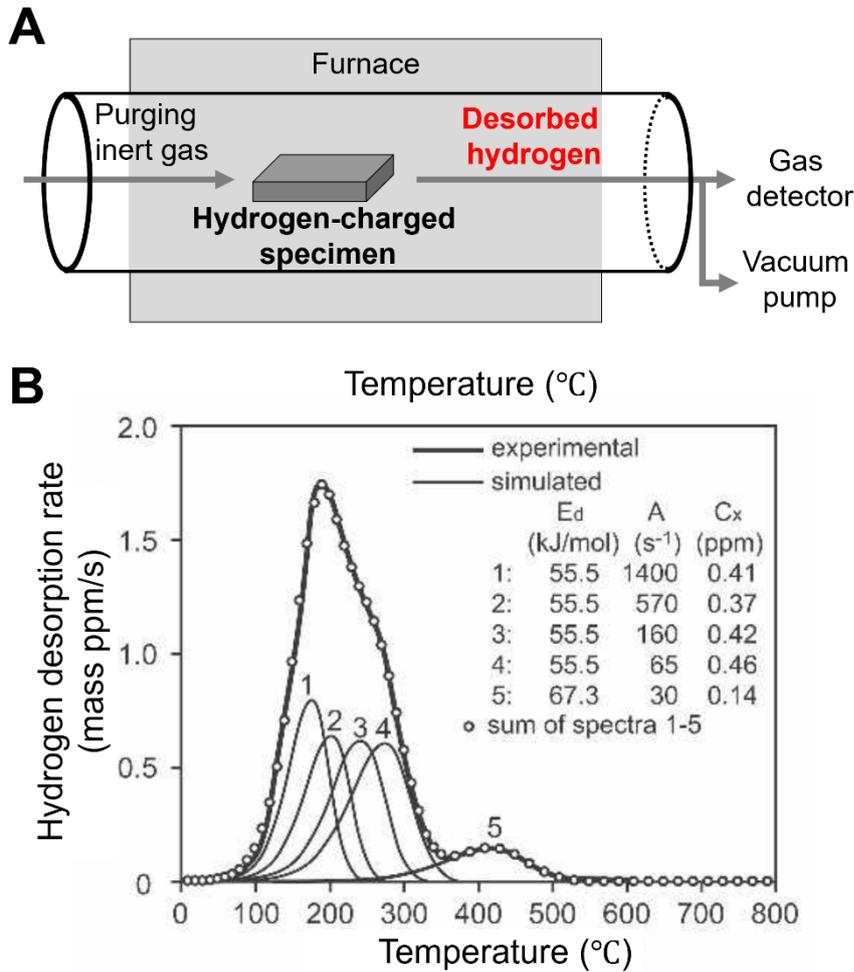

**Figure 21. Thermal desorption analysis for the study of hydrogen trapping.** (A) Thermal desorption spectrometer setup with inert gas purging. (B) Example TDA data and peak deconvolution from a martensitic steel quenched from the austenite temperature (experimental and simulated datasets) [262].

The time-to-test, the environment, and the specimen dimensions are the most significant variables for hydrogen quantification by TDA [156]. **Figure 22**A shows that the amount of hydrogen loss from iron specimens over time (where hydrogen absorption is endothermic) is greater if the specimens were held in vacuum than in argon, **Figure 22**B shows that smaller specimens retain less hydrogen, and that there is fast hydrogen loss in vacuum, consistent with findings in [268]. These issues explain some of inconsistency of hydrogen measurements between research groups and raise issues that need to be considered for quantitative hydrogen analyses on microscopic specimens where hydrogen egress is rapid, such as specimens prepared for microscopy experiments. Recent developments in low-temperature TDA (L-TDA, or cryogenic TDA as C-TDA) addresses some of the challenges in retaining hydrogen that desorbs at room temperature or below [185,186].

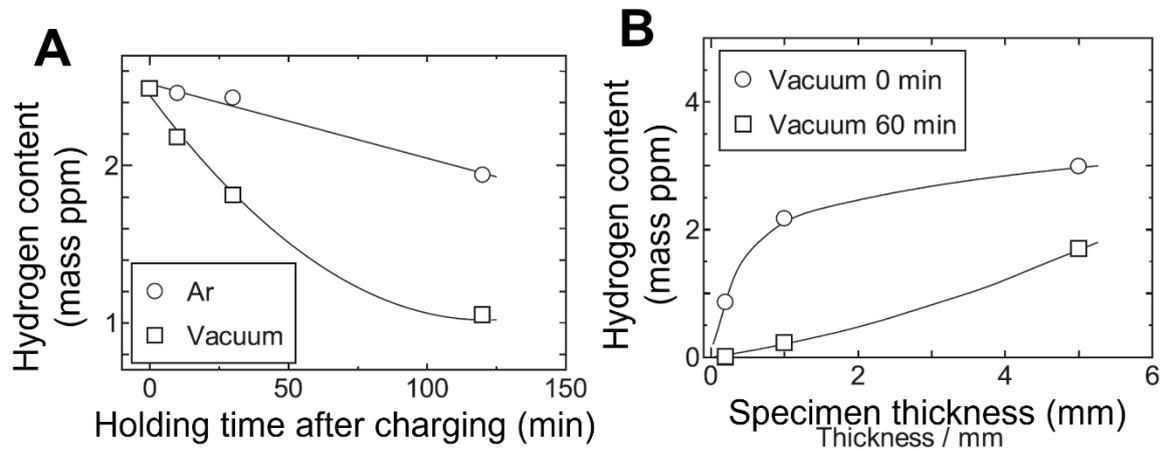

**Figure 22. The influence of TDA experimental parameters.** Hydrogen in iron samples as a function of (A) holding time after hydrogen charging in both argon and vacuum environments and (B) thickness after exposure to vacuum for two different durations [156].

*3.2.3. Hydrogen permeation*

A hydrogen permeation setup and the steady-state hydrogen concentration profile through the specimen are illustrated in **Figure 23**A and B [187]. The specimen is generally in the form of an approximately 1 mm thick sheet (L in **Figure 23**B), and is the working electrode of both electrolytic cells in **Figure 23**A. Hydrogen is generated at the specimen surface in the left-hand cell by cathodic hydrogen charging either at a constant potential ($\Delta V_1$), as noted in the figure, or a constant current density. The applied potential generates a hydrogen fugacity that is designated as $f_{H2}$ in **Figure 23**B, which establishes an equilibrium hydrogen content ($C_H$) in the specimen that can be evaluated via Sieverts' law, i.e. Equation (9), and can be evaluated from the steady state hydrogen flux through the specimen. The hydrogen diffuses through the metallic specimen from the hydrogen-charging side to the exit side where the hydrogen flux is measured as a current as the hydrogen is oxidized in the oxidation cell by the application of a suitable potential ($\Delta V_2$) to the working electrode. The technique is sensitive to small hydrogen fluxes through the specimen as it is possible to reliably measure a

small current [158,180,272,273].

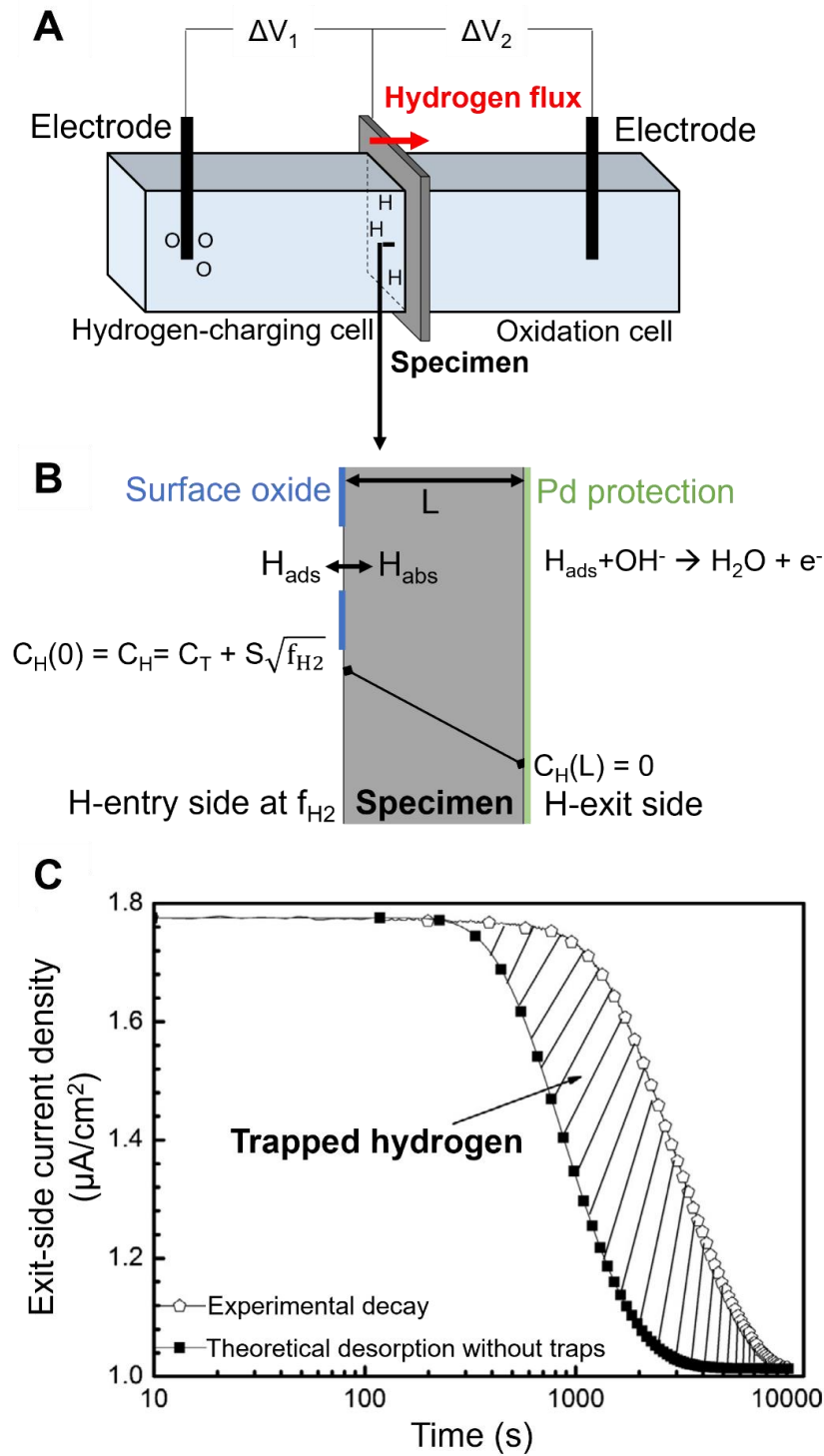

**Figure 23. Hydrogen permeation.** (A) Schematic of the hydrogen permeation experiment. (B) Activities at the specimen surface. (C) Example hydrogen trapping measurement by a decay transient (from -1.2 V to -1.1 V). A 3.5NiCrMoV medium strength steel and 0.1 NaOH charging solution were used. (C) is reproduced from [273].

A typical permeation experiment uses an inert gas such as nitrogen sparged into the hydrogen entry cell to prevent the evolution of surface gas bubbles that can affect the active area for the reaction. De-aeration of the solution in the oxidation cell is used to remove the current associated with the oxidation of dissolved oxygen in the solution. The exit surface of the specimen is typically coated with Pd (or Ni) to prevent oxidation. Generally, there is a surface oxide on any metal specimen (regardless of the specimen preparation method), and long-term pre-charging has been shown to be required for hydrogen charging of steels at moderate hydrogen fugacities to reduce the surface oxide to a stable condition and avoid irreproducible hydrogen permeation transients. The hydrogen concentration is linear across the specimen toward the exit side where $C_H = 0$ under steady state conditions.

Any alloy specimen has reversible hydrogen traps, as indicated in **Figure 13**, and the decrease of the hydrogen charging potential ($\Delta V_1$) decreases the entry-side hydrogen fugacity leading to a decay transient measured at the exit side. **Figure 23**C shows how the hydrogen decay transient can be used to estimate the amount of reversibly trapped hydrogen by comparing it with the simulated data for a trap-free specimen. The trapping energy can also be obtained by adjusting the energetic factors such as temperature [273].

*3.2.4. Microscopy techniques*

In recent years, microscopy has been used to correlate hydrogen and microstructural features. Microscopic characterization of hydrogen is challenging for four reasons. First, hydrogen is mobile in metallic specimens and can desorb from the specimen rapidly. Second, microscopic specimens have generally small dimensions and total hydrogen egress is faster from a specimen with small dimensions. Third, conventional electron microscopes typically cannot detect hydrogen, due to the low atomic mass of hydrogen and low interactivity with the incident electron beam [274–277]. For this reason, electron-based techniques are not generally useful for the characterization of hydrogen unless it is chemically bonded in stable phases. Fourth, residual hydrogen gas is present in the vacuum system of ion beam or electron microscopes [278–280], so the source of any detected hydrogen is uncertain [281,282]. The techniques available for the characterization of hydrogen trapping are summarized in a comprehensive review by Koyama et al. [86]. Here we introduce these techniques in three groups according to their mode of hydrogen detection.

Firstly, neutron beams can provide a contrast between hydrogen isotopes and other elements based on their distinct scattering cross-sections [283]. Neutron imaging can be used to map the distribution of hydrogen with sub-millimeter precision by comparing specimens with and without hydrogen [284,285]. Neutron imaging is non-destructive and can also provide reconstructed 3-D data.

Secondly, spatial distribution of hydrogen can be mapped by using spatially resolved mass spectroscopy techniques such as secondary ion mass spectroscopy (SIMS) [286] and atom probe tomography (APT) [29,86]. The methods utilize ion bombardment and field evaporation, respectively, to remove and detect atoms from the specimen, including hydrogen, allowing the creation of atom maps at the micro- and nano-scale. Both techniques are sensitive to light elements, but the analysis of hydrogen is complicated by the presence of hydrogen as a residual gas in their vacuum chambers. Uncertainty over the origin of the hydrogen detected can be circumvented by using deuterium ($^2$H or D), the second most abundant hydrogen isotope. Deuterium has a larger atomic mass that is easily differentiated from the primary isotope of hydrogen (protium, $^1$H) by mass spectrometers, and can be used during charging (if charging with hydrogen is a part of the experiment) [87,198,235,286–293]. The different masses and hence slightly different diffusivities of $^1$H and $^2$H [208] do not normally significantly influence the distribution of hydrogen within samples at ambient conditions [294]. Hydrogen loss is a particular problem because of the small specimen size for SIMS and APT [293]. For APT, hydrogen loss has been mitigated using cryogenic sample transfer to inhibit hydrogen desorption, which led to cryo-APT [87,198,288,289,291,292,295–297]. Cryogenic transfer/analysis for SIMS is also possible and has been demonstrated for a ferritic steel [298], but is not yet available for the NanoSIMS. Even without cryogenic capabilities, NanoSIMS has been successfully used to analyze hydrogen charged samples for a range of metallographic systems [235,290,299–301].

Thirdly, hydrogen trapping can also be characterized by decorating the sample with a chemical or physical agent that can interact with the hydrogen and can represent the location of hydrogen. The techniques using silver decoration and an integrated microscope with good spatial resolution was designated. The silver ions on the surface of a hydrogen-charged specimen are reduced by the hydrogen that is desorbed from the specimen bulk according to the following reaction:

$$Ag^+ + H_{ads} \rightarrow Ag + H^+ \qquad (12)$$

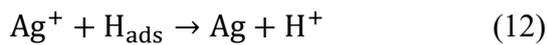

The $Ag^+$ ion can be deposited on the surface using an Ag-containing chemical such as AgBr. The reduced metallic Ag atoms reveal the hydrogen-rich regions. The hydrogen microprint technique (HMT) operates similarly to the silver decoration method, relying on the reduction and deposition of silver atoms [302–304]. The main distinction lies in the use of an Ag(CN)2 aqueous solution, immersing the specimen surface for silver decoration, in contrast to the approach using an AgBr emulsion. Another similar method decorates the sample surface with hydride-forming materials such as palladium to induce a localized change in conductivity, which can be measured using a scanning Kelvin probe (SKP) which detects hydrogen at the sample surface. The technique in combination with high-resolution scanning probe microscope is called scanning Kelvin probe force microscopy (SKPFM) [305–307]. These techniques require a surface that is oxide-free prior to the application of the hydrogen-sensitive decoration. The temporal and spatial resolutions of these techniques were summarized by Koyama et al. as shown in **Figure 24** [86].

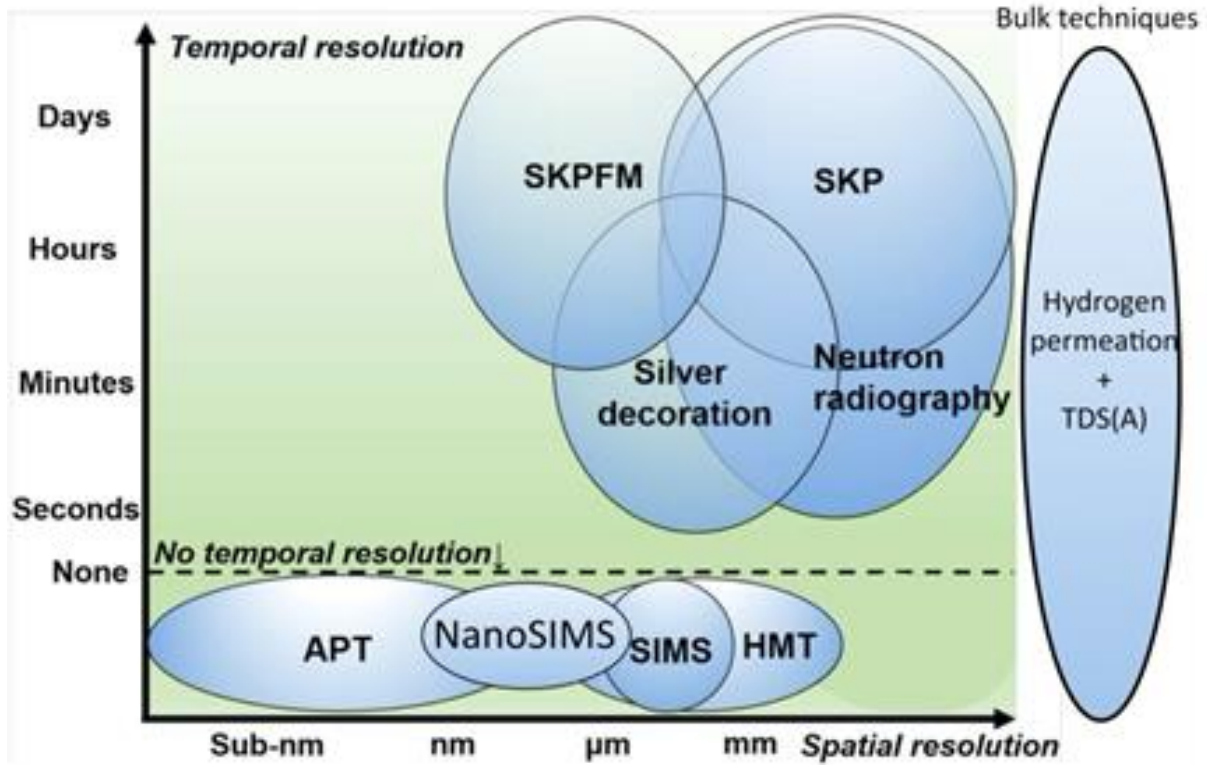

**Figure 24. Overview of spatial and temporal resolutions of common hydrogen mapping techniques.** SKP(FM): scanning Kelvin probe (force microscopy). APT: atom probe tomography. SIMS: secondary ion mass spectroscopy. HMT: hydrogen microprint technique. TDS(A): thermal desorption spectroscopy (analysis). Reproduced from [86].

*3.3. Characterization of hydrogen trapping*

The different microstructural features that act as hydrogen traps were illustrated in **Figure 17**. **Figure 25** and **Table 1** illustrate the hydrogen trapping energies of common traps ($-E_b$) in a BCC iron matrix. The scatter is attributed to both the trap structure and measurement details [188]. **Figure 25** indicates that lattice defects (dislocations and GBs) and precipitates with coherent interfaces have a median trapping energy less than 50 kJ/mol (often defined as the reversibility limit); whereas less coherent precipitates have higher hydrogen trapping energies. Many early studies (particularly those before 1990 [287,288]) did not consider the contribution of the substructures of the trapping features, e.g. vacancies in carbides. It may now be prudent to revisit the trapping properties of some of these trap types now that better techniques are available to measure the hydrogen within different phases and at different microstructural features.

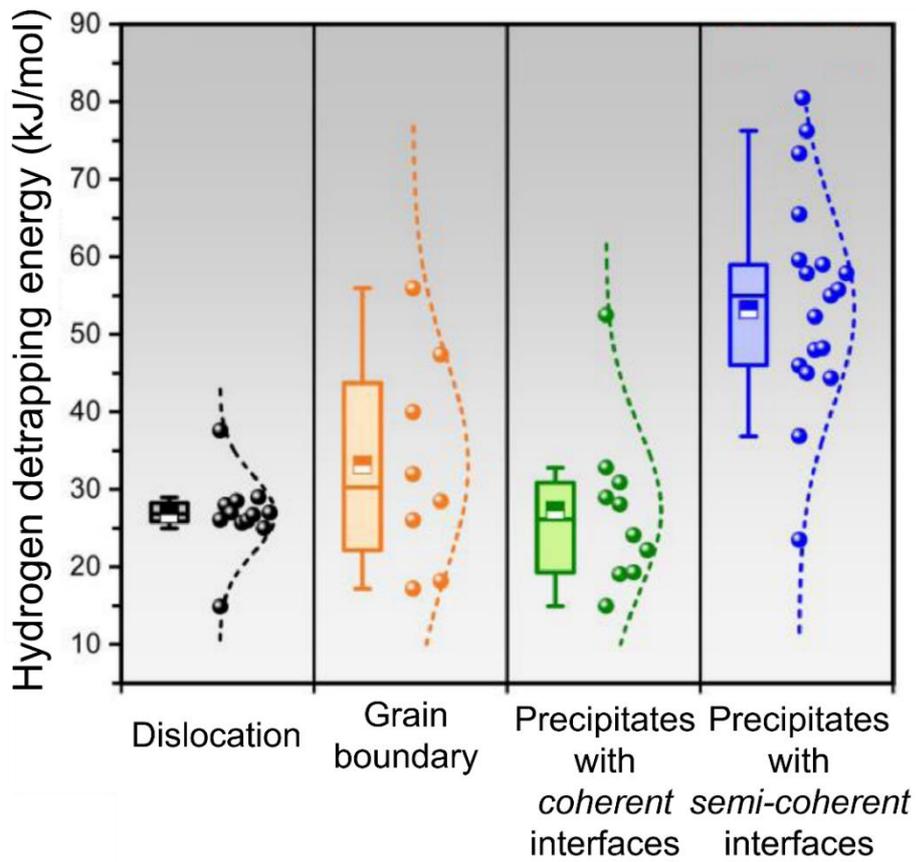

**Figure 25. Comparison of literature data for hydrogen trapping energies (-$E_b$) for common features in the BCC iron matrix**. Reversibility of hydrogen trapping is defined as 50 kJ/mol. Reproduced from [233]

**Table 1. Trapping energies of common hydrogen traps in the BCC iron matrix**

| Trap type | Binding energy (-$E_b$, kJ/mol) | Characterization technique | Reference |
|---|---|---|---|
| **Solute elements** | | | |
| Si, Cr, Mn, Co and Mo | Approx. 0 | First Principle | [225] |
| C (interstitial) | 9 | First Principle | [225] |
| N (interstitial) | 13 | Magnetic Relaxation | [310] |
| Nb | -7* | First Principle | [225] |
| Ti | -8* | First Principle | [225] |
| Mg | -15* | First Principle | [225] |
| Sc | -20* | First Principle | [225] |
| Y | -25* | First Principle | [225] |
| **Crystal defects** | | | |
| Single vacancy | 24-78 | First Principle, Diffusion Analysis | [225,311–314] |
| Micro-voids | 40 | TDA | [309] |
| Dislocations (bulk) | 60 | Diffusion Analysis | [315] |
| Screw dislocations | 26 | First Principle | [204] |
| Dislocation strain field | 12-27 | Diffusion Analysis, TDA | [194,316,317] |
| Grain boundary | 9-49 | Mechanical Analysis, TDA | [317–320] |
| Prior austenite grain boundary | 47 | Permeation | [321] |
| **Second phases** | | | |
| Incoherent TiC | 60–129 | Permeation, TDA | [262,308,322–324] |
| Semi-coherent TiC | 48 | TDA | [262] |
| Incoherent $V_4C_3$ | 40 | TDA | [325] |
| Semi-coherent $V_4C_3$ | 25-28 | TDA | [326,327] |
| Incoherent NbC | 55-60 | TDA | [328] |
| Semi-coherent NbC | 28-56 | TDA | [328,329] |
| Coherent $Mo_2C$ | 11-34 | TDA | [330–332] |
| Cementite/α interface | 11-18 | Permeation, TDA | [317,333,334] |
| ε carbide | 12-65 | Permeation, TDA | [335,336] |
| ε copper | 27 | TDA | [320] |
| Dispersed oxide | 45 | Permeation, First Principle | [337,338] |
| MnS interface | 64 | TDA | [309,316] |
| Austenite/Ferrite interface | 44 | Permeation | [339] |
| Iron oxide interface | 43-62 | TDA | [340] |
| $Al_2O_3$ interface | 71 | TDA | [341] |

*Negative value means repulsive

*3.3.1. Solute elements*

Hydrogen trapping at a solute element in a metal matrix is generally insignificant compared to other microstructural traps. Counts et al. used first-principle techniques to study the interaction energies of various solutes with hydrogen in a BCC iron matrix [225]. The interstitial carbon solute atom was found to prefer to occupy the octahedral site, unlike hydrogen which prefers the tetrahedral site [190]. The interaction between the two atoms, 9 kJ/mol as noted in **Table 1**, is insignificant compared to the diffusion energy in the BCC matrix (4.5-5.5 kJ/mol as per [194]). The substitutional solute atoms were found to be either inert or repulsive to nearby hydrogen atoms (**Table 1**), and the presence of these elements does not significantly influence the interaction between a hydrogen atom and a vacancy, meaning that the hydrogen trapping at the atomic scale is dominated by the vacancy itself, regardless of the presence of solute elements. These findings were supported by later research [342].

*3.3.2. Crystal defects*

*Hydrogen-enhanced strain-induced vacancies*

Part 3.3.2 describes hydrogen-enhanced strain-induced vacancies as a mechanism for embrittlement. Hydrogen stabilizes vacancies in many metals and alloys [32,47,121,129,222,225,343–348]. Nagumo and Takai [47] reviewed this phenomenon and described how hydrogen-enhanced vacancy formation and subsequent vacancy clustering can lead to the formation of microvoids, which then coalescence, leading to hydrogen-induced crack propagation by ductile micro-void coalescence. An extremely high hydrogen charging fugacity (GPa level) can lead to the formation of stable hydrogen-vacancy clusters in an abnormally high concentration (up to 30 atomic percent), referred to as 'superabundant vacancies' by Fukai [349].

Hydrogen can also be trapped by dislocations, similar to the Cottrell atmosphere formed by other interstitial solute atoms such as carbon in steel [121,129,130]. Hydrogen trapping at dislocations in an untempered martensitic steel specimen was directly observed by Chen et al., using cryo-APT [87]. **Figure 26**A and B are bright- and dark-field TEM images of the dislocation-dense martensitic structure. The APT reconstruction in **Figure 26**C contains carbon isosurfaces that mark the position of dislocations. **Figure 26**D shows the colocalization of atoms of carbon (blue) and deuterium (red) at the dislocations, providing evidence supporting the HELP model [30,31]. The opposite effect - hydrogen-*reduced* dislocation mobility was found by Xie et al. [350] in an aluminum alloy using TEM with in-situ mechanical loading in a hydrogen-containing environment. The authors suggested that this could be caused by an additional interplay with vacancies. Hydrogen enhanced mobility occurs for both edge and screw dislocations [351]. Calculations indicate that the dislocation core interacts with hydrogen more strongly than the elastic field of the dislocation and that the increase in core radii and decrease in core energy leads to the reduction of dislocation line energy by H [352]

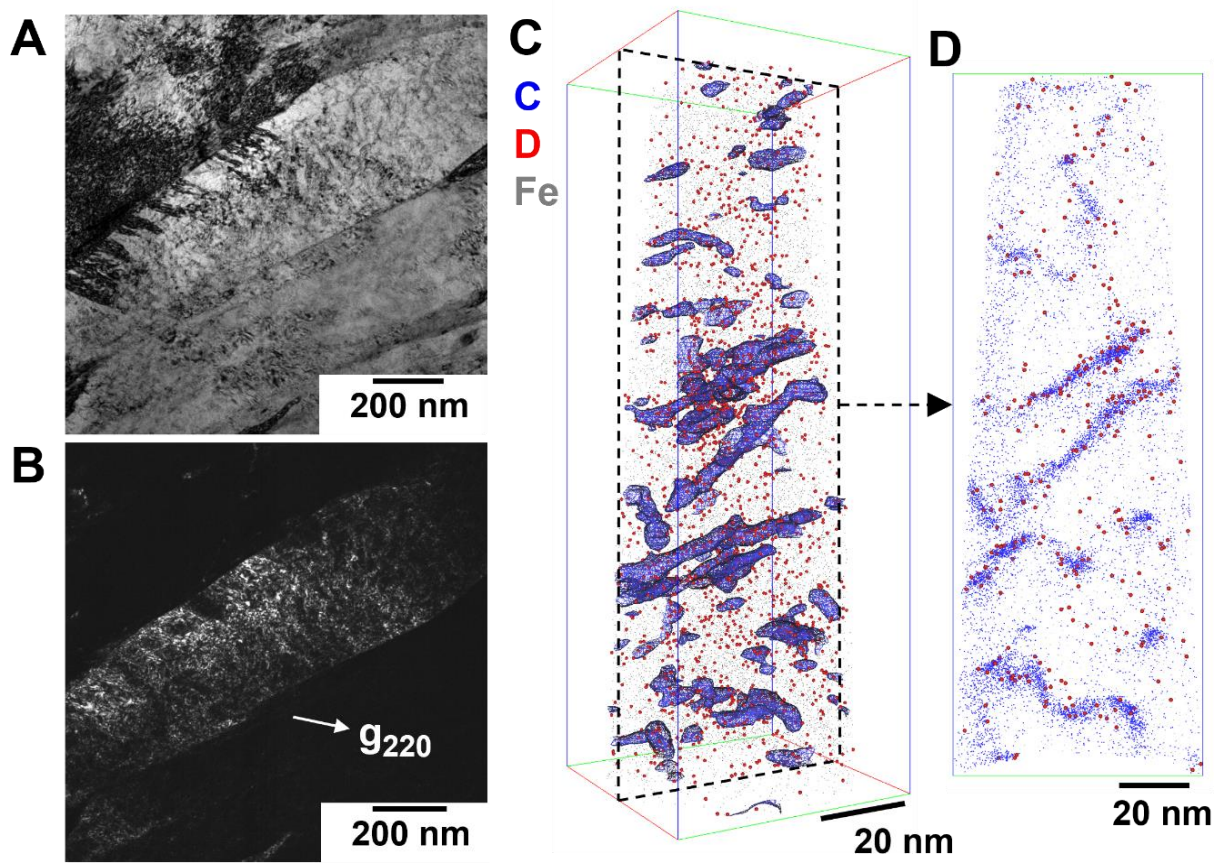

**Figure 26. Dislocation hydrogen trapping in steel.** (A) Bright-field image of dislocations in a martensitic steel, (B) dark-field image of (A). (C) 3-D APT map of carbon (blue), deuterium (red) as a marker for hydrogen, and iron (grey) with isoconcentration surfaces (blue surfaces) highlighting the location of dislocations represented by clustered carbon. (D) 2-D slice from the marked region in (A) showing the coincident locations of carbon (blue, i.e., dislocations) and deuterium (red). Reproduced from [87]

In addition to acting as a hydrogen trap, a mobile dislocation can also drag hydrogen. Hydrogen transport by dislocations can dominate hydrogen re-distribution [353–355]. NanoSIMS analysis of a 625+ Ni alloy showed deuterium enrichment at dislocation slip bands after a crack growth test with in-situ charging. However, when the alloy was charged but not strained, there was no deuterium localization, indicating deuterium transport by mobile dislocations [290]. Dislocation movement induced by hydrogen diffusion has also been recently recorded [356].

Since GBs interfere with dislocation glide, hydrogen transport by dislocations can lead to hydrogen accumulation at GBs, which may promote hydrogen-induced decohesion and subsequent intergranular fracture [357,358]. This relationship between strain and hydrogen distribution may also explain the strain rate dependence of HE, as discussed in Section 2.1.4. A high strain rate results in high-speed dislocations that cannot transport hydrogen, reducing hydrogen drag and causing less embrittlement. A low strain rate produces slower-moving dislocations that transport hydrogen, contributing to HE.

Grain boundaries can also trap hydrogen, which has been observed by APT [87]. Compared to vacancies and dislocations, hydrogen behavior is more complex at GBs, and depends on the crystal structure of the host matrix and the geometric structure of the GB. In BCC metals, GBs generally act as trapping sites since the hydrogen diffusivity at GBs is lower than the lattice diffusivity. In BCC iron, hydrogen generally reduces the GB cohesion, which leads to fracture at a sufficiently high hydrogen concentration [265]. Prior-austenite GBs are particularly susceptible [123]. In FCC metals, where lattice diffusion of hydrogen is slower, GBs can act as hydrogen traps and offer a rapid diffusion path for hydrogen, depending on the relative diffusivity in the FCC lattice and along the particular GBs [360]. In austenitic steel, $\Sigma 5$ and $\Sigma 9$ GBs trap hydrogen, but not $\Sigma 3$ GBs [361]. HE in FCC nickel alloys generally leads to intergranular fracture. The effect of hydrogen on grain boundary behavior remains difficult to predict. The exact interplay of GBs with hydrogen is a topic deserving of more emphasis in future research [113,362].

Twin boundaries (TB) are more prevalent in FCC metals such as nickel. Using SKPFM and TDA, Koyama et al. concluded that TB trapping is only slightly stronger than other reversible traps such as dislocations [33,363]. Aboura et al. used NanoSIMS to identify deuterium localization at twin boundaries in a deuterium-embrittled nickel alloy [290]. **Figure 27**A is a secondary electron image from this study that contains a region with a deformation-induced twin boundary (TB, green arrow). Protium ($^1$H, the reference hydrogen background in this experiment), and deuterium ($^2$H, real signal) were mapped, as shown in **Figure 27**B and C, respectively. The ratio between the logarithmic concentrations of $^2$H and the logarithmic concentration of $^1$H from **Figure 27**B and C, respectively, then led to **Figure 27**D which shows a bright line (high $^2$H intensity) along the identified TB, showing deuterium trapped at the TB. Although trapping at TB has been observed, the exact role of TB for bulk HE, i.e., whether hydrogen at TB facilitates or mitigates HE, is still a topic of discussion [33,363].

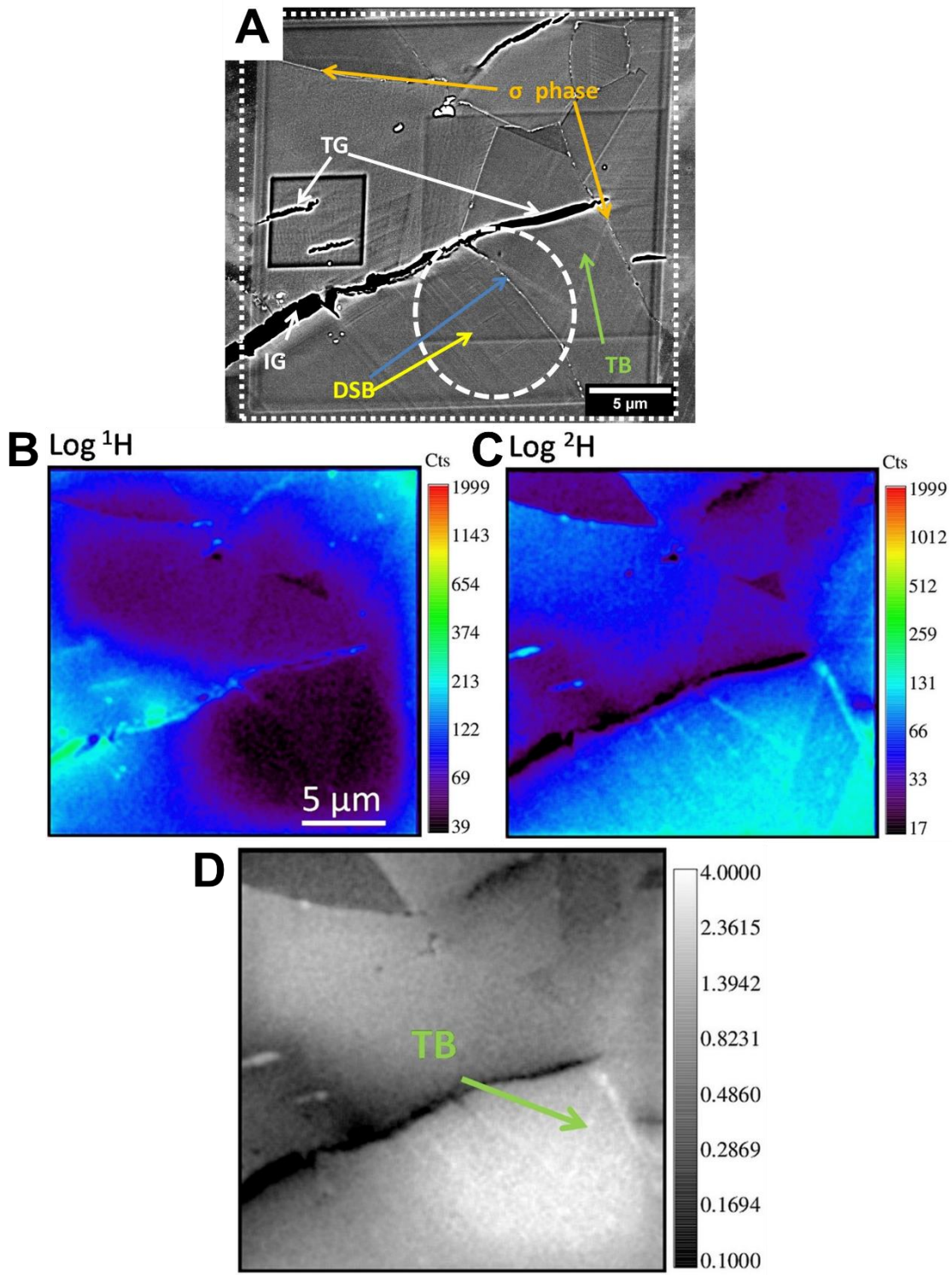

**Figure 27. NanoSIMS evidence of twin boundary hydrogen trapping in a nickel-based alloy.** (A) Secondary electron image of a region of interest (ROI) at the proximity of a crack in a deuterium-embrittled alloy 625+ sample. The ROI contains the secondary σ phase, dislocation slip band (DSB), transgranular and intergranular cracks (TG and IG, respectively), and deformation twin boundary (TB). (B) and (C) are the logarithmic secondary ion intensity maps of protium ($^1H$) and deuterium ($^2H$), respectively. (D) is the ratio of the logarithmic $^2H$ to logarithmic $^1H$ ratio map which shows the concentration of deuterium at TB (green arrowed region). Reproduced from [290]

*3.3.3. Precipitates / second phases*

Precipitates and/or second phase particles in a solid solution matrix can act as benign hydrogen traps and can significantly increase HE resistance by reducing the available hydrogen [17,28,49,297,364]. In ferritic steels, fine precipitates of the carbides of transition metals, e.g., Ti, V, Nb, and Mo, can be produced with a high number density, with tunable hydrogen trap strengths, which are thought to be controlled by their interface coherencies [365–369].

DFT has been used to investigate the hydrogen trapping mechanism of carbides with a rock-salt (NaCl) structure (**Figure 28**A), such as TiC (TEM image in **Figure 28**B), $V_4C_3$, and NbC [218,370]. **Figure 28**C shows the calculated hydrogen solution energy around the interface between a TiC particle (right-hand side) in ferrite (left-hand side) [370]. The authors found that i) the interface of the TiC particle is an effective trap; ii) a carbon vacancy inside the TiC bulk is the strongest trap, however it is difficult for hydrogen to reach this trap because of the high energy barrier; iii) this energy barrier can be lowered in the presence of a high number of carbon vacancies that form a channel for hydrogen penetration into the TiC bulk (black broken line); and iv) the presence of extra hydrogen atoms in a trap can significantly reduce the de-trapping energy and facilitate the penetration of hydrogen, which is likely the case for high-fugacity/electrolytic hydrogen charging (red broken line). **Figure 28**D shows the trapping energies of a range of metal and non-metal vacancies in common FCC precipitates evaluated using DFT [234]. This calculation suggests that carbon vacancies are stronger hydrogen traps than metal vacancies.

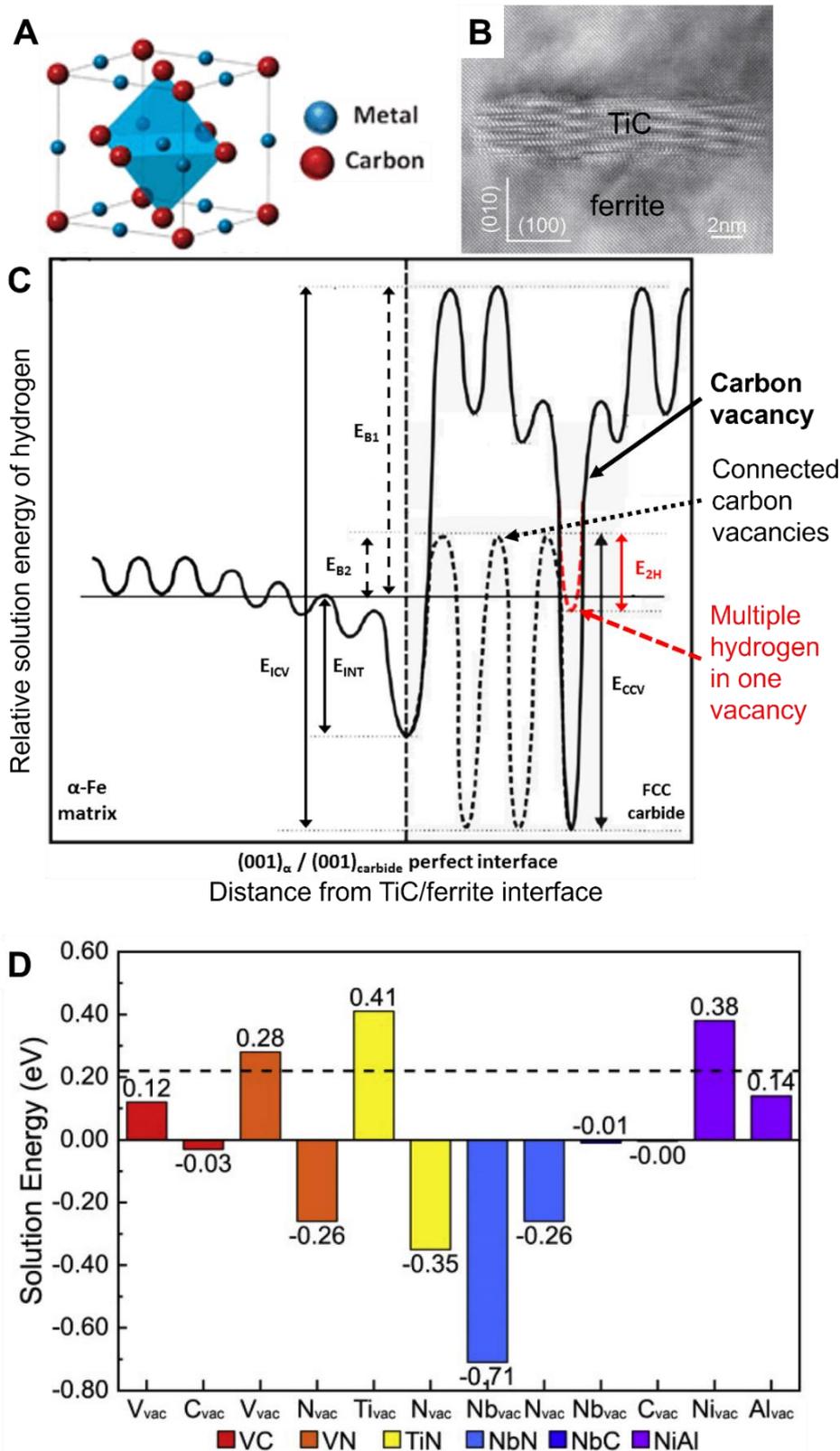

**Figure 28. Density function theory (DFT) calculations for hydrogen trapping in TiC**. (A) Schematic of the FCC carbide structure. (B) Typical TEM image of TiC in ferrite [262]. (C) Solution energy of hydrogen around a TiC precipitate with different configurations[370]. (D) Solution energy of hydrogen in metal and non-metal vacancies for common FCC precipitates [234]. It is important to note that the hydrogen potential differs between (C) and (D) and, as such, cannot be directly compared

Atom Probe Tomography (APT) has been used to study hydrogen trapping at fine FCC carbide precipitates [198,288,289,291,292]. Takahashi and co-workers pioneered the use of gaseous hydrogen charging and cryo-APT to study hydrogen trapping at TiC and $V_4C_3$ as shown in **Figure 29**A and B, respectively, confirming that these carbides act as hydrogen traps in steels. [198,289]. Later, Chen et al. [196] used cryoAPT to investigate the hydrogen trapping mechanism of a deuterium-charged V-Mo-mixed carbide. They concluded that hydrogen was located within the precipitates using a statistical approach to normalize and superimpose the concentration profiles across all of the measured particles, as shown in **Figure 29**C.

The V-Mo carbides studied by Chen et al. had more carbon vacancies than those studied by Takahashi et al. due to the presence of molybdenum as a substitutional metal element at vanadium sites [238]. Chen et al.'s finding of interior trapping of hydrogen in carbon-vacancy-containing carbide was consistent with Di Stefano et al.'s numerical prediction about hydrogen penetration in the presence of abundant carbon vacancies and charged hydrogen [370]. More recently, Chen et al. used the same protocol to investigate hydrogen trapping in a well-annealed, spherical, incoherent NbC, which should have few carbon vacancies. **Figure 29**D shows that hydrogen was trapped at the NbC/ferrite interface, consistent with the computational prediction that the interface acts as a hydrogen trap [233]. Similarly, SKPFM studies showed trapping at the interface for large, incoherent TiC particles in a ferritic matrix [371]. Hydrogen trapping in this type of FCC carbide is therefore related to both the interface and the presence of defects within the precipitate.

In addition to transition metal carbides, cementite and epsilon (ε) carbide in steels also act as hydrogen traps as demonstrated in [372–374] and [287], respectively. These carbides are believed to primarily trap hydrogen in the carbon vacancies at the interface, although high-resolution observations are required for verification.

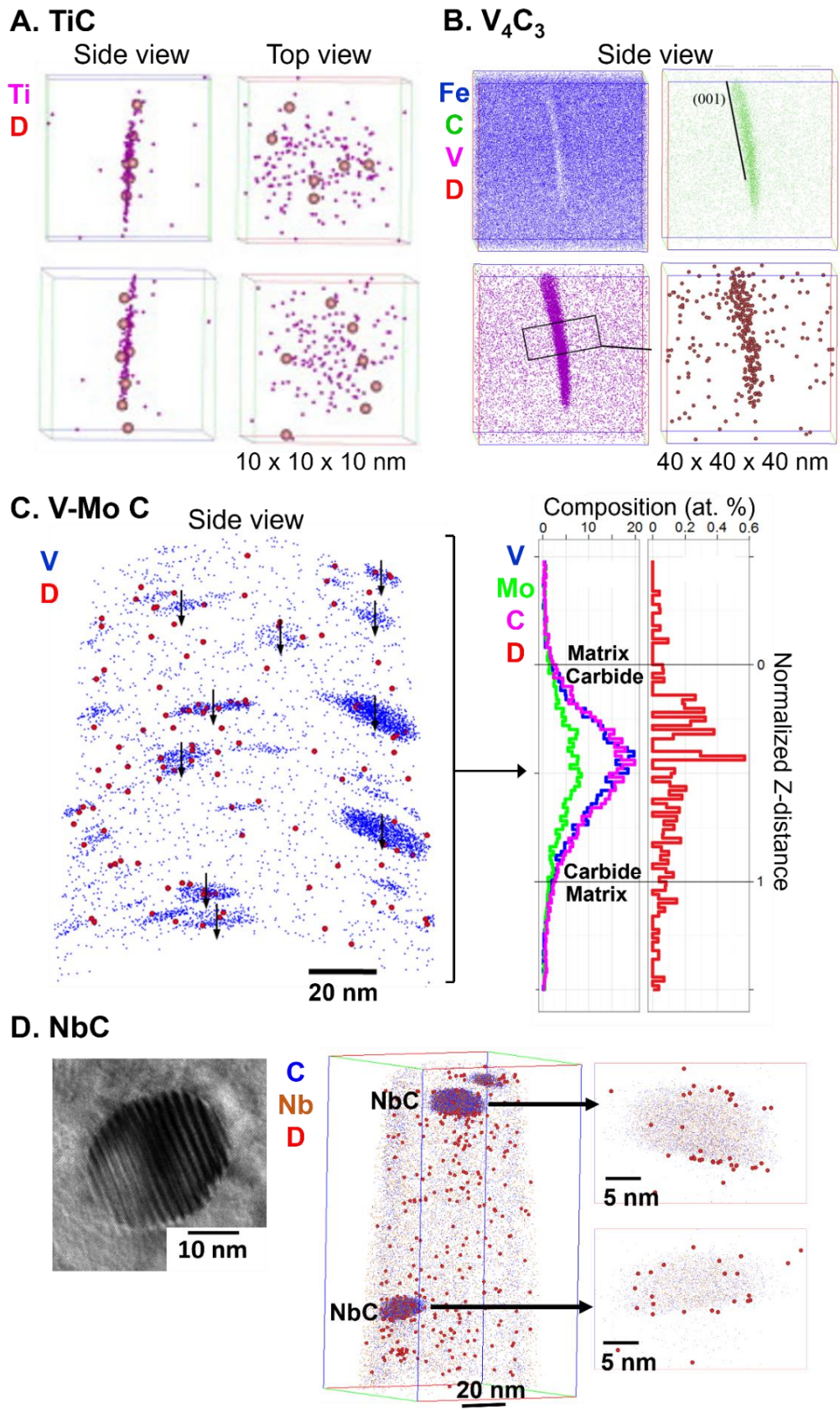

**Figure 29. APT characterization of hydrogen trapping by face-centered cubic/rocksalt-**

**structure carbides.** (A) Hydrogen (deuterium) in TiC [289]. (B) Result for $V_4C_3$ [198]. (C) Result for V-Mo carbide from [196]. (D) Result for NbC from [87].

Apart from carbides, metallic particles (such as copper) have also been identified as hydrogen traps [320,375–377], although the exact trapping sites for these particles has not been demonstrated. In oxide dispersion strengthened (ODS) steels, simulations suggest that hydrogen trapping occurs through oxygen-vacancy pairs at the interface of oxide particles [378]. Some sulfide inclusions such as MnS can in theory also trap hydrogen strongly, and such inclusions are known to cause local HEDE at their incoherent interfaces where the hydrogen concentration leads to fracture initiation [140,379–383].

Retained austenite, an FCC phase in the BCC iron matrix, is also an important hydrogen trap. Hydrogen has a much lower diffusivity and a much higher solubility in austenite than in ferrite or martensite. [384–390]. APT visualizations of hydrogen in retained austenite are shown in **Figure 30** [385]. In some advanced high-strength steels, retained austenite in the microstructure can undergo an austenite-to-martensite phase transformation during deformation. This transformation-induced plasticity (TRIP) can significantly increase ductility [391]. However, this transformation releases any hydrogen dissolved in the austenite (as the austenite transforms) into the freshly formed martensite, which can be severely embrittled by the hydrogen [28,386,388,392]. Therefore, the use of austenitic phases as hydrogen traps requires caution that any subsequent loading or thermal environment does not lead to detrimental hydrogen release [393,394].

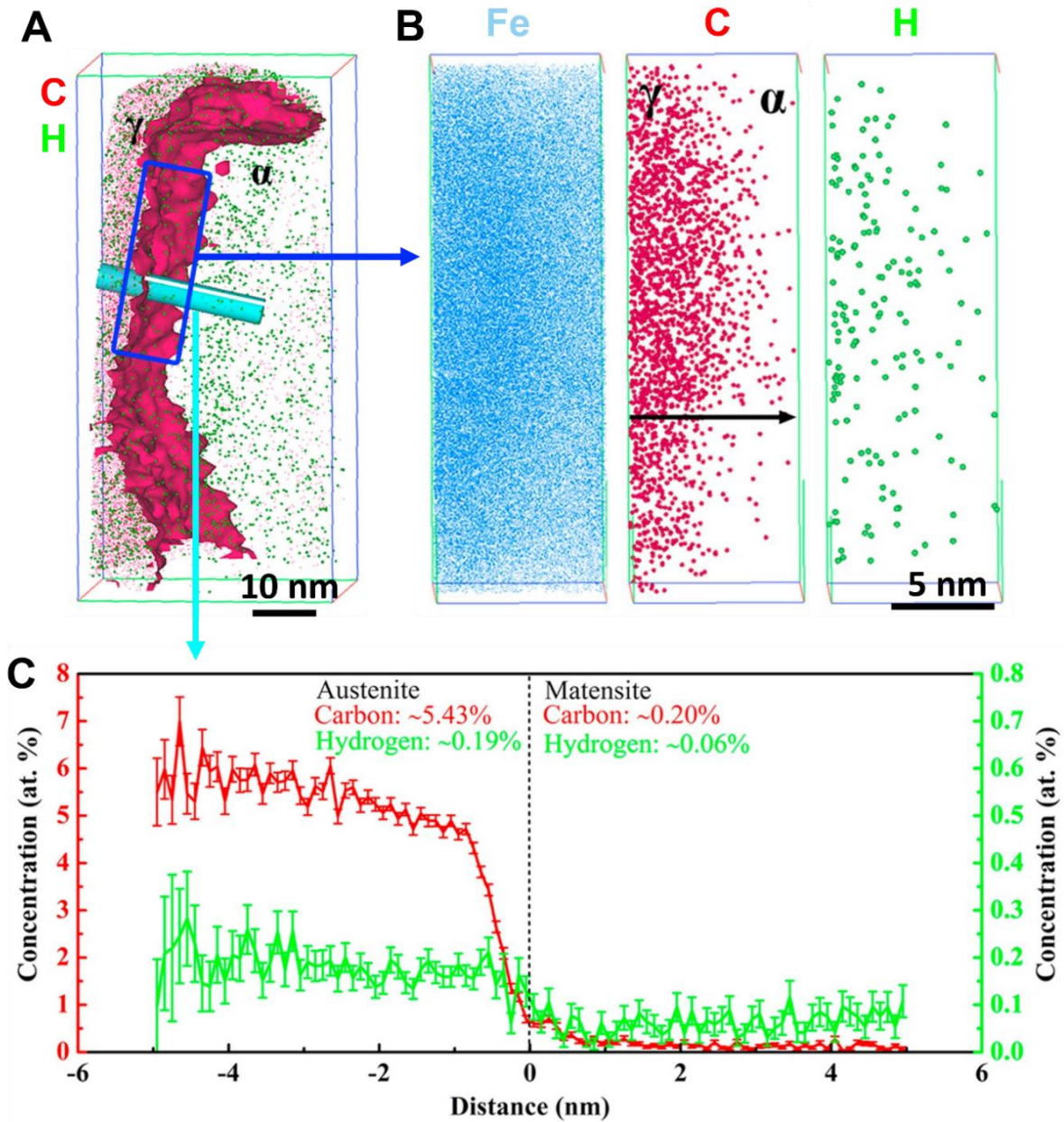

**Figure 30. Hydrogen trapping in retained austenite in ferrite.** (A) 3-D APT map with austenite-ferrite interface highlighted by a carbon isoconcentration surface. (B) Atom maps of iron, carbon, and hydrogen ($^1$H, not deuterium) from the highlighted region in (A). (C) Atomic concentration across the interface from cylindrical the region highlighted in (A) [360].

In hydride-forming metals such as titanium and zirconium, some second phase particles have also been found to trap hydrogen [236,395–397]. APT observations of a Ti-Mo alloy, **Figure 31**A, revealed that hydrogen can be trapped at the interface of the Mo-rich second phase [396]. NanoSIMS observations have also shown the hydrogen trapping at the Fe- and Cr-rich phases in a Zr-Fe-Cr alloy, **Figure 31**B [235]. These hydrogen traps can play an essential role in the redistribution of hydrogen upon thermal cycling in the nuclear applications in which these Zr alloys are used, which can cause precipitation and re-orientation of hydrides and can lead to delayed hydride cracking [98].

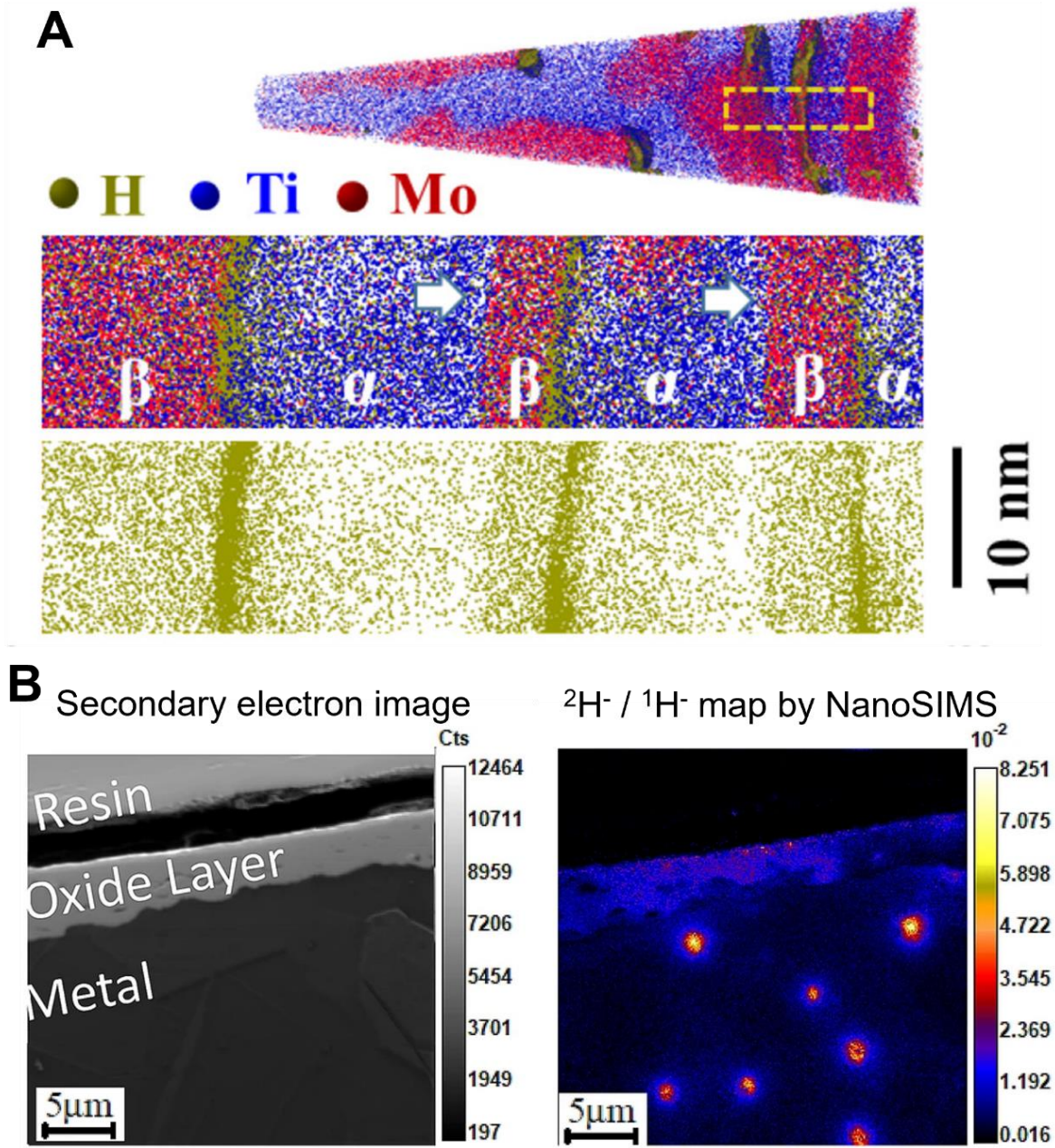

**Figure 31. Second phase hydrogen trapping in titanium and zirconium alloys.** (A) APT atom maps of the primary α-Ti phase (blue), secondary β-Ti-Mo phase (red), and hydrogen (dark yellow) [396]. (B) NanoSIMS mapping of hydrogen trapping in Zr-Fe/Cr phases in zirconium alloys [235]

4. **Microstructural engineering of hydrogen traps for embrittlement mitigation**

For a closed system with a finite amount of hydrogen (i.e., IHE), the addition of microstructural hydrogen traps is expected to be effective for HE mitigation [28]. Where hydrogen supply is infinite (i.e., EHE), traps are not generally thought to be viable strategy for mitigation, as the traps saturate in an open system with a permanent source of hydrogen [398], however modelling has suggested that trapping can reduce the critical hydrogen concentration in open systems under cyclic loading [200]. In this section, we review the role of hydrogen traps for the mitigation of HE.

Exploiting trapping behavior in the design of embrittlement-resistant alloys requires a better understanding of the relationship between microstructural hydrogen trapping and macroscopic HE mitigation. First, we reiterate the key considerations proposed by Pressouyre [28,139,399] to define whether a trap is 'good' or 'bad' for reducing HE susceptibility in a particular application:

1. What is the form and the possible amount of the hydrogen supply? i.e., is hydrogen supply internal or environmental, continuous or intermittent, infinite or finite?
2. Is the material subjected to elastic or plastic deformation in the presence of hydrogen? i.e., what is the interplay of dislocations and plasticity with hydrogen?
3. What microstructural sites are most susceptible to HE? i.e., does the HE-induced fracture occur transgranularly, intergranularly or at a second phase interface?
4. What is the strength of the benign hydrogen traps in a material compared to that of the HE-susceptible features at the temperature of service? i.e., do the benign traps compete with the features that also attract hydrogen, but contribute to HE, such as dislocations or MnS interfaces? Are the traps reversible?

### 4.1. *Grain boundaries and dislocations*

A finer grain size and a lower dislocation density increase EHE resistance (**Figure 8**) [91]. Grain refinement is effective in a range of alloy systems including ferritic steels [400,401], martensitic steels (prior-austenite grain size) [123,402], austenitic steels [403–408], nickel [409], and FCC high-entropy alloys [410,411]. The benefit is attributed to increased hydrogen trapping due to more GB areas and, for the case where hydrogen supply is finite, less hydrogen per unit GB area [412]. A graded grain structure with fine-grains at the surface can also provide better HE resistance than a homogeneous grain structure [413], while retaining a high strength-ductility combination [414].

Dislocations are generally considered to be traps that are unfavorable for HE mitigation [415]. In an exception, it was recently reported that for severe deformation of pure iron by cold work, above a certain dislocation density, the HE drops slightly due to the trapping effect of dislocation cell walls [416]. **Figure 32**A shows that HE increased as cold work increased from 0% to 70% and decreased thereafter (red arrow). TEM revealed that the 80% cold work specimen has a substructure of dislocation cells, as shown in **Figure 32**D. The authors attribute this decrease of HE susceptibility an effect similar to grain refinement, i.e., increasing the density of trapping sites and limiting the availability of hydrogen within the dislocation cells, as illustrated in **Figure 32**E.

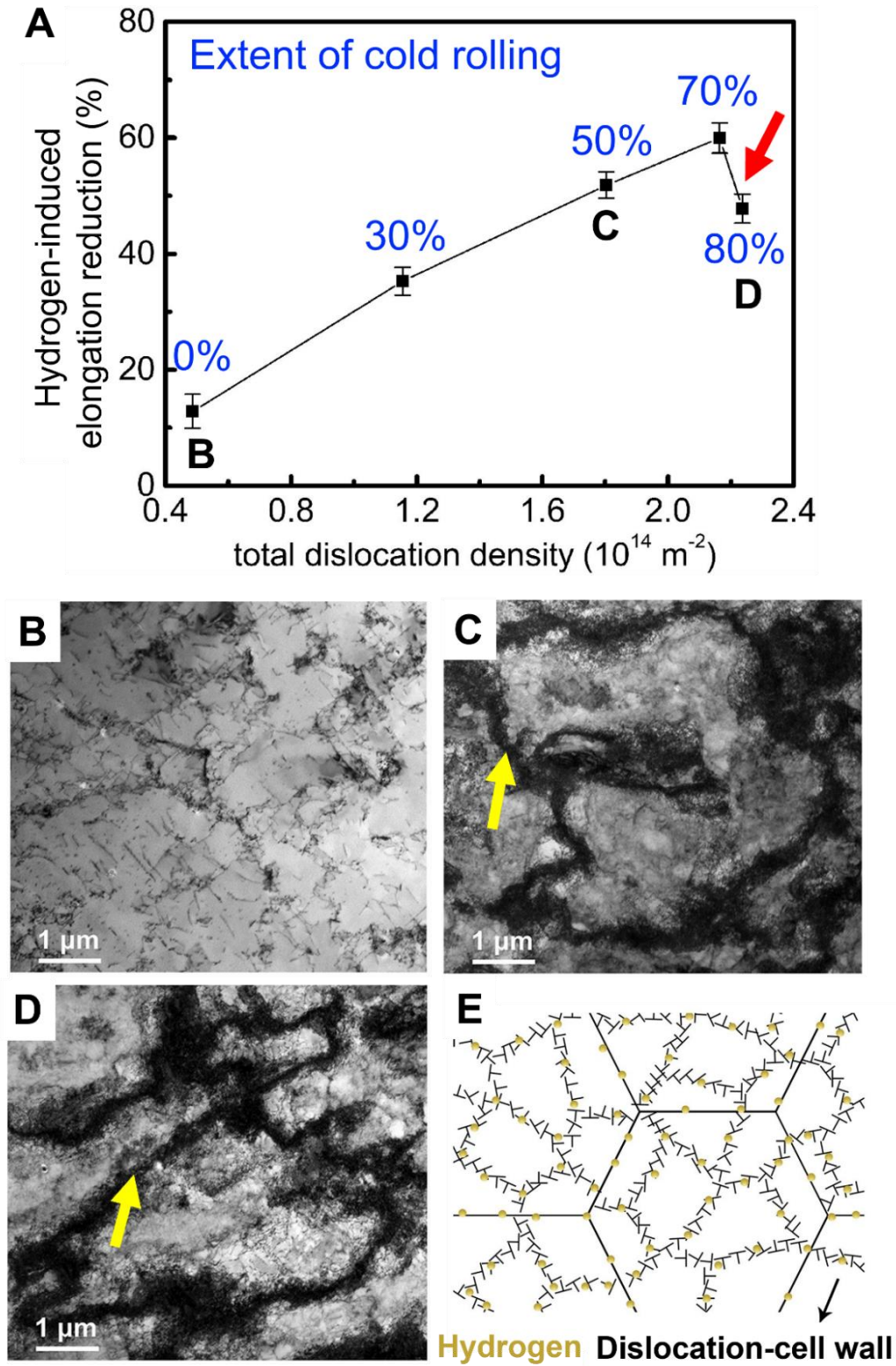

**Figure 32. Effect of dislocation cells on internal hydrogen embrittlement (IHE) susceptibility.** (A) IHE susceptibility (defined by elongation loss) as a function of dislocation density, relating to the extent of cold work. TEM micrographs of dislocation structures in (B) undeformed, (C) 50%-deformed, and (D) 80%-deformed specimens. (E) Schematic illustration of the dislocation cells trapping hydrogen to increase the HE resistance. Reproduced from [416]

It has also been possible to fabricate alloys with desired GB orientations to mitigate GB fracture, an approach known as 'grain boundary engineering' [417,418]. The efficacy of GB engineering for reducing HE susceptibility in FCC alloys was demonstrated by Bechtle et al. [124]. Lower HE susceptibility was measured in the Ni alloy specimens with more low-energy boundaries (such as twin boundaries). Similar results were later measured for other FCC alloy systems [419–422]. The concept is that a higher density of 'special' GBs that have higher coherency with adjacent grains (i.e., $\Sigma$ number $\leq 29$, such as $\Sigma 3$, $\Sigma 5$, etc.) lead to a lower tendency to trap hydrogen atoms. This GB engineering approach has only been demonstrated for FCC alloys, which are more susceptible to hydrogen-induced intergranular failure than BCC alloys.

The interaction of hydrogen and other species segregated at GBs is still largely unknown, particularly the effects on hydrogen trapping and HE [215]. Segregation also varies from boundary to boundary and depends on the boundary structure. Correlative TEM-APT [423] allows measurement of both structure and elemental segregation at GBs. Future studies may also use NanoSIMS, cryo-APT and/or mechanical testing to enable a better understanding of the relationship between elemental segregation, hydrogen trapping and HE.

*4.2. Hydrogen trapping at particles*

Fine precipitates have drawn the most attention in the area of hydrogen trapping for HE mitigation because they have the potential to be incorporated into microstructures at significant number densities and hence provide a high trapping capacity [305,328,424–428]. They are also known to have a concomitant hardening effect [429,430]. Research in this area has mainly focused on BCC steels with carbide or nitride precipitates based on transition metals (e.g., titanium, vanadium, molybdenum, chromium, and niobium).

The use of microstructural traps to inhibit hydrogen assisted fatigue was theoretically described by Fernandez-Sousa et al. [200] for conditions where the loading cycle was significantly faster than the time required to deliver hydrogen to the process zone by either bulk diffusion or dislocation transport. Efficacy for IHE was experimentally demonstrated by Verbeken, Depover, and their co-workers, using a protocol combining TDA, SSRT, heat treatment (to control the number density of precipitates) and TEM for carbonitride characterization [328,424–428]. **Figure 33** [424] is an example of these results. **Figure 33**A and B are example TEM images of carbide precipitates. TDA provided the hydrogen content of each specimen (**Figure 33**C) for charging conditions in which the hydrogen reached equilibrium (hydrogen saturation) after a range of heat treatment conditions that led to various carbide densities. The hydrogen capacity was related to the number of carbides, increasing with the number of traps. All specimens were susceptible to HE when saturated with hydrogen. However, specimens with similar hydrogen contents that contain precipitates (here, TiC) displayed lower HE susceptibility (defined by ductility loss). In **Figure 33**C, the as-quenched (As-Q) specimen of Alloy C was saturated at 7 wt. ppm hydrogen, but the quenched-and-tempered (Q&T) Alloy C with more traps had a capacity of 12 wt. ppm (highlighted in red). These specimens were heat-treated for 10 min, 2 hours, and 1 hour to create specimens that contained little TiC, a low number density of TiC (over-aged), and a high number density of TiC (peak-aged). Specimens were deliberately charged with 7 wt. ppm hydrogen for mechanical testing, in order to match the hydrogen content in the differently aged specimens. The tensile strengths and elongations for charged and uncharged specimens in each condition are shown in **Figure 33** D. Those charged with 7 wt. ppm hydrogen are marked 'charged*'. For ease of comparison, each heat treatment condition was labeled the uncharged (blue) and 7 wt. ppm charged (red) stress-strain curve. The samples containing TiC (**Figure 33**E and F) had significantly less loss in ductility when charged with hydrogen. In addition to the number density of carbides, the size of carbides can significantly impact hydrogen embrittlement resistance, particularly with variations in tempering time. In this work, titanium carbides were found to enlarge and lose coherency during the tempering process. Consequently, the corresponding carbides grow excessively large, and their interface with the matrix becomes too incoherent to adequately trap hydrogen from an electrochemical source.

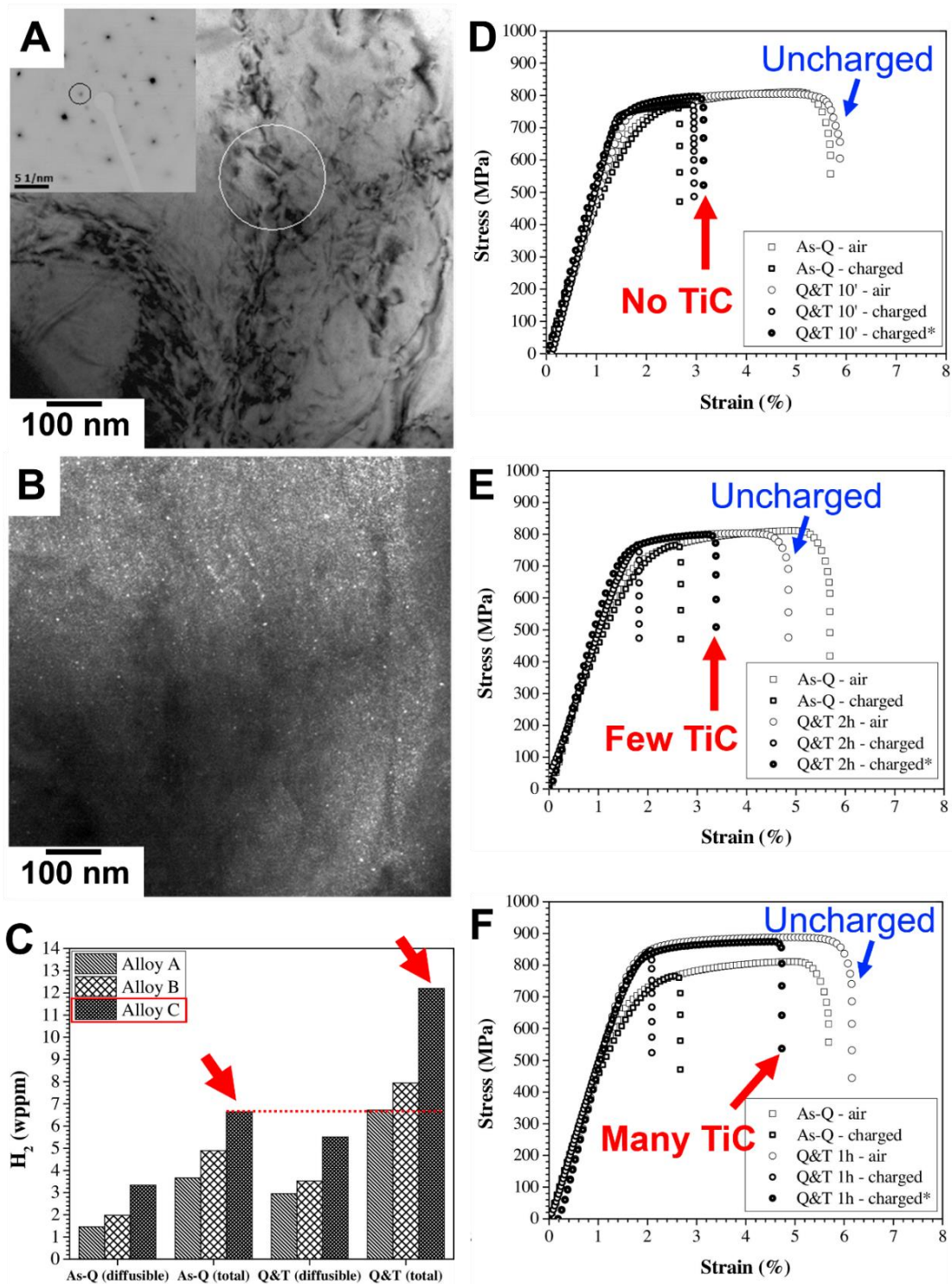

**Figure 33. Efficacy of TiC for internal hydrogen embrittlement (IHE) resistance.** (A)

Bright-field TEM micrograph of a specimen with high density of TiC (Alloy C charged for 1h, corresponding to the data shown in (F)). (B) Dark-field TEM micrograph of (A) from the second-phase diffraction spot indicated inset in (A). The bright spots are TiC precipitates. (C) Saturated hydrogen content in various tested materials showing the greater capacity for H in the sample with precipitates (Alloy C, Q&T). (D), (E), and (F) highlight the tensile properties of uncharged specimens (blue arrows) and the specimens charged with 7% hydrogen (red arrows) from the materials containing negligible, low, and high numbers, of TiC. Reproduced and adapted from [424]

Verbeken, Depover, and their co-workers studied steels containing vanadium, molybdenum, and chromium carbides [425–427], and also confirmed the beneficial role of these carbides for mitigating IHE, consistent with an earlier study of vanadium-alloyed steels [431]. A comparison of the mitigation efficacy of Mo, V, and Cr carbides found that, for similar steel strength, Mo and V carbides were more effective than Cr carbides for HE mitigation [432]. In alloys with the same composition, the microstructures with finer carbides and higher trap density trapped hydrogen more effectively [433]. For NbCs, small well-dispersed NbCs reduced HE susceptibility [434], although undissolved NbC precipitates remaining after austenization did not contribute measurably to HE resistance [435]. More recently, tantalum carbide precipitate (TaCs) was found to have a similar effect on HE mitigation [436,437].

For martensitic steels, Ti and Mo carbides were confirmed to have a positive effect on HE mitigation [432,438]. To further understand HE mechanisms in martensitic steels that contain oxide and carbide inclusions, Feaugas, Guedes, and their co-workers combined SSRT testing on notched samples with in-situ hydrogen permeation, TDA, fractography, and finite element modeling (FEM) [140] (**Figure 34**). They concluded that the trapped hydrogen leads to ductility loss, but still ductile fracture, whereas the presence of both mobile and trapped hydrogen caused greater ductility loss and brittle fracture. They compared plate tensile martensitic samples with both V- and U-sharped notches under three conditions: i) uncharged, ii) hydrogen-charged and then room-temperature desorbed (leaving only trapped hydrogen in the specimens) and iii) in-situ hydrogen-charged with a high hydrogen flux near the surface. **Figure 34**A shows that the specimens with trapped hydrogen (green curve) were slightly less ductile than the uncharged specimens (green), but the specimens with both mobile hydrogen and trapped hydrogen (blue curves, labelled H flux) were much more susceptible to HE. Other studies showed little HE in hydrogen pre-charged then desorbed specimens with a low hydrogen content at the surface [439–441]. **Figure 34**B and C show, respectively, the ductile (dimpled) fracture surface of the uncharged specimen and the specimen with trapped hydrogen. Both specimens contained oxides at the base of the dimples, but the specimens with trapped hydrogen also had dimples associated with small carbide inclusions (**Figure 34**D). The specimens with high hydrogen flux at the surface displayed brittle fracture with the HE-characteristic quasi-cleavage (**Figure 34**E). These observations suggest that inclusions and carbides, which are hydrogen traps, can enhance ductile fracture, but also be responsible for the nucleation of cracks if diffusible hydrogen is present.

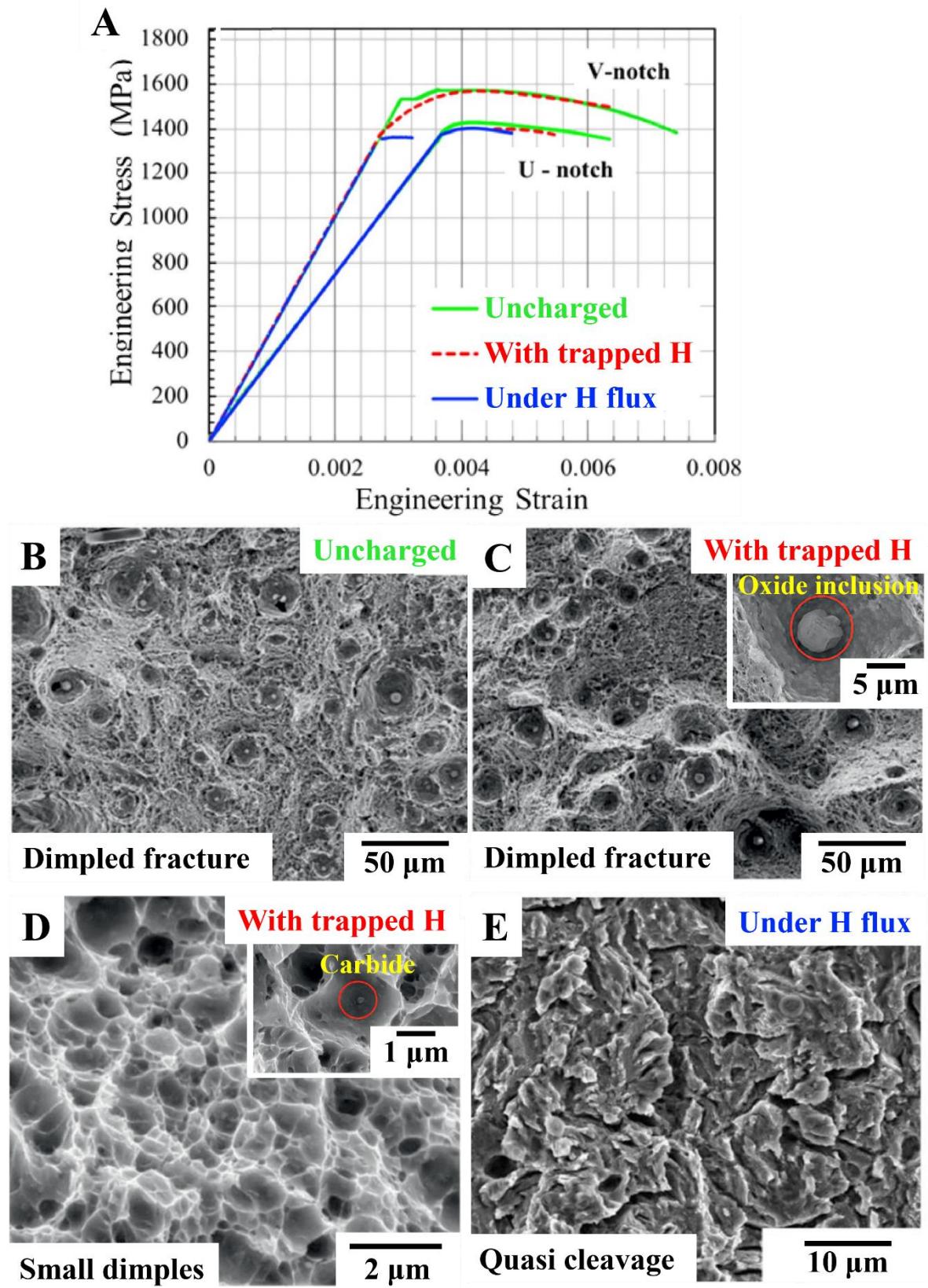

**Figure 34. The role of trapped and diffusible hydrogen in HE.** (A) SSRT results from a Fe-0.3C-0.4Si-0.5Mn-1.0Cr-0.8Mo-0.05V-0.04Nb (wt. %) martensitic steel with small amounts (less than 0.04 wt.%) of Al, S, Cu, Co and Ca, which is austenitized, quenched, and tempered at 710 °C for 30 min. Both V- and U-shaped notches were used in plate tensile samples. The ductility decreased in the following order: specimen without hydrogen (green curves), specimens with trapped hydrogen (red curves), and specimens with diffusible and trapped hydrogen (blue curves). SEM fracture surface observations of (B) uncharged specimen with dimpled features, (C) specimen with trapped hydrogen, also with dimpled features showing oxide inclusions at the centers of the dimples, (D) the specimen with trapped hydrogen showing additional carbide inclusions at the centers of the dimples. (E) Quasi-cleavage for specimens with both trapped and mobile hydrogen. Note the scale difference among the micrographs. Reproduced from [140]

*4.3. Microstructural engineering using second phases*

Another strategy is to take advantage of the difference in hydrogen diffusivity in different phases [442]. Because hydrogen diffuses more slowly in austenite than in ferrite, percolated, three-dimensionally interconnected austenite (**Figure 35**A) leads to lower hydrogen permeability than an open structure (**Figure 35**B) or one in which the austenite is not fully interconnected (**Figure 35**C). This is demonstrated in **Figure 35**D, which shows the hydrogen diffusivity for steel samples with different fractions of austenite and different levels of interconnectivity. The first three datapoints (circles) show that the hydrogen diffusivity decreases with an increasing fraction of retained austenite in bainite, which can be associated with the degree of connectivity of the austenite. However, the hydrogen diffusivity of a conventional equiaxed ferritic-austenitic microstructure (square data point) is much higher, even though the fraction of austenite is higher, which can be explained by the fact that, unlike for bainite, the austenite is not interconnected (percolated) in this microstructure.

Utilizing austenite as a hydrogen barrier, Sun et al. examined the HE-induced deformation modes of two medium-Mn steels, one austenite-based (percolated) and one ferrite-based (not percolated) schematically illustrated in **Figure 35**E and F, respectively [443]. Ductile deformation behavior of the ferrite-based microstructure was attributed to hydrogen-enhanced local plastic flow. The austenite-based microstructure failed by interfacial decohesion attributed to hydrogen accumulation at phase boundaries and a subsequent strain-induced martensitic transformation that is particularly susceptible to HE.

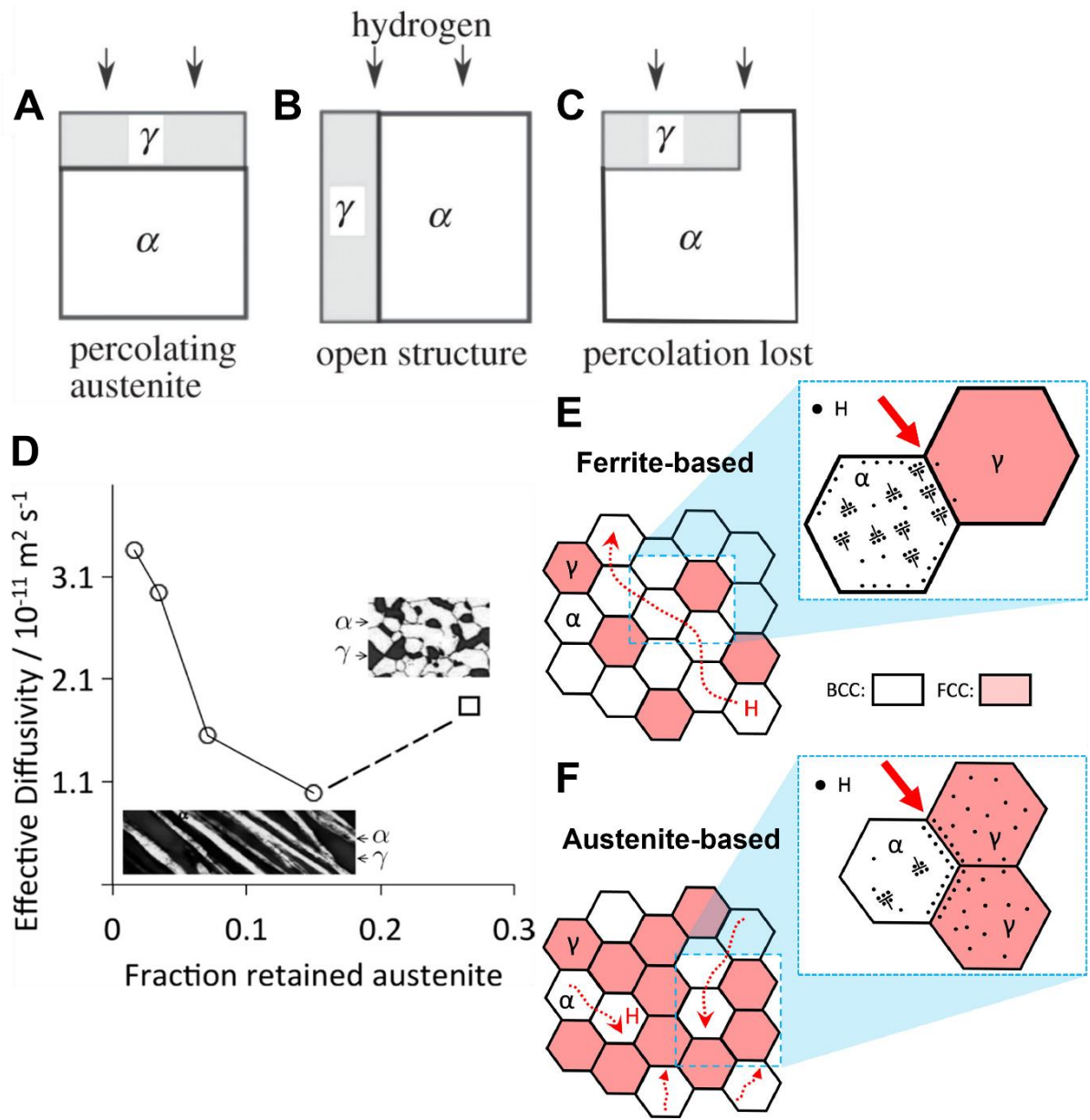

**Figure 35. Hydrogen percolation in austenite-containing steels.** (A), (B), and (C) are schematic illustrations of a microstructure with percolating austenite, open structure, or a loss of percolation, respectively, with respect to the hydrogen ingress from the top. (D) shows hydrogen diffusivity as a function of the fraction of retained austenite for percolated austenite in a bainitic ferrite matrix (circular data points) compared to a conventional austenite-ferrite microstructure (square data point). (A)-(D) are reproduced from [342]. (E) and (F) are schematic illustrations of the hydrogen diffusion and distribution of a steel matrix with disconnected (E) and interconnected (F) austenite, reproduced from [443].

To further optimize the design of a HE-resistant microstructure, Sun et al. took advantage of typically undesirable chemical heterogeneity within the austenite grains in a medium-Mn steel [444] to achieve superior embrittlement resistance. **Figure 36**A and B show the austenite-ferrite dual phase microstructure with an uneven Mn distribution by electron backscattering diffraction (EBSD) and electron energy-dispersive X-ray spectroscopy (EDX), respectively. The mechanism is shown schematically in **Figure 36**C and D. Mn enrichment increases the stability of the austenite and suppresses the phase transformation to martensite (**Figure 36**C). The Mn-stabilized austenite remains ductile, blunts the crack tip, and ultimately reduces the rate of crack propagation (**Figure 36**D). **Figure 36**E shows that HE resistance (represented by the elongation of hydrogen pre-charged specimens) of the specimen with Mn heterogeneity was improved by factor of two.

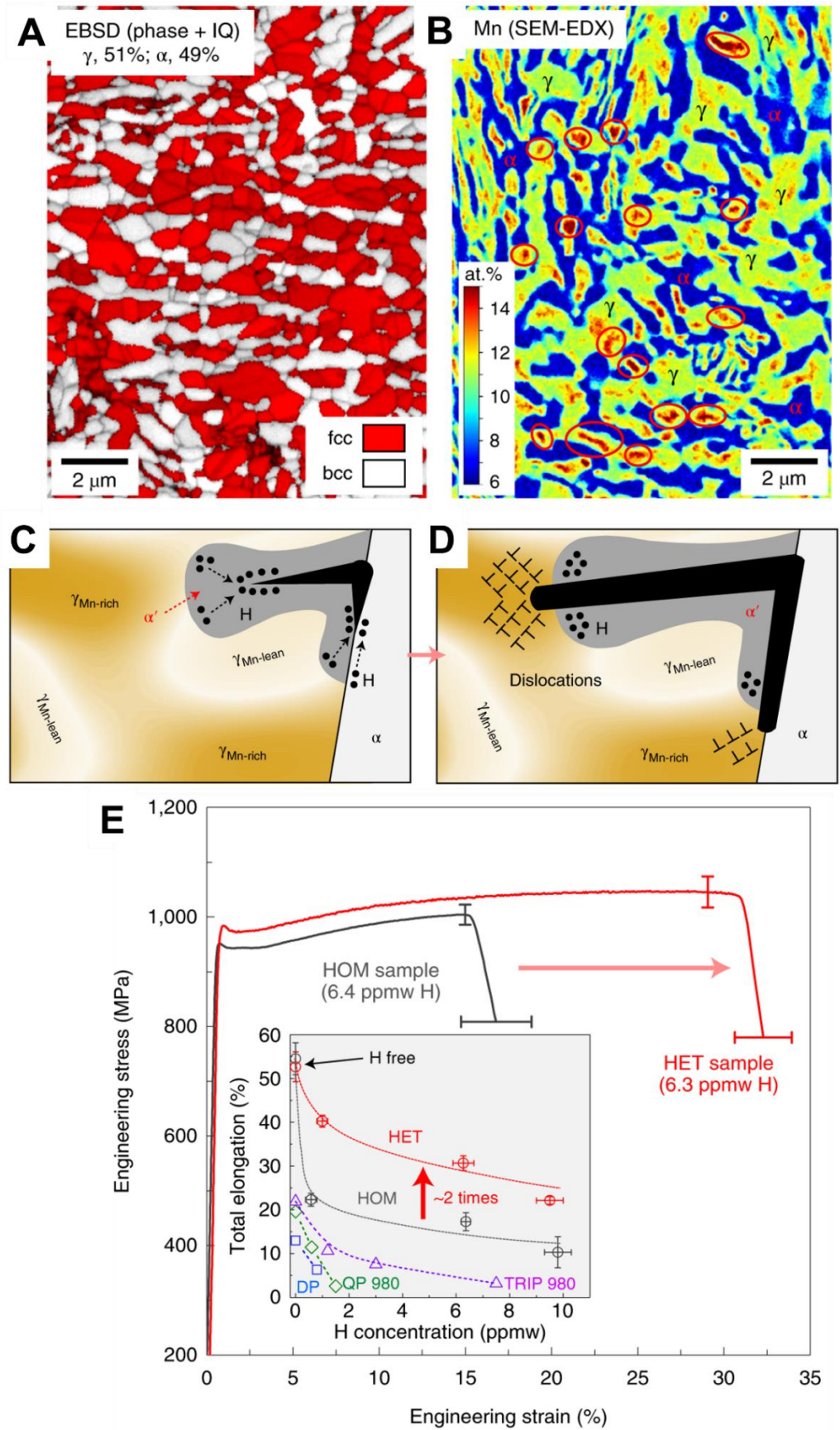

**Figure 36. Chemical heterogeneity engineering for HE resistance.** (A) and (B) are EBSD and EDX maps showing the austenitic phase distribution and Mn heterogeneity, respectively.

(C) and (D) are schematic illustrations of the crack propagation in an austenite grain with Mn heterogeneity, where the Mn-rich regions are less prone to strain-induced martensitic transformation. (E) shows the HE performance of specimens with (red) and without (black) Mn heterogeneity, suggesting that the Mn heterogeneity led to higher HE resistance. Reproduced from [444].

## 5. Summary and outlook

We have examined the variables, the causes, and the mechanisms of HE, introduced the principle of hydrogen trapping, described the techniques that have advanced the understanding of hydrogen trapping, and discussed several microstructural designs that can provide increased HE resistance. Much work has been devoted to understanding hydrogen trapping and many studies indicate that the incorporation of suitable traps can decrease EHE when the transport of hydrogen to the process zone is limited, but there is still much research required to further advance our understanding of HE and hydrogen trapping for the development of HE-resistant alloys.

Hydrogen interactions with engineering alloys are multiscale. An understanding requires an understanding of the interaction between hydrogen atoms and the atomic scale microstructural features (grain boundaries, phases, precipitates, solute atoms, solute clusters, interfaces, as well as the multi-scale features that control plasticity and failure (vacancies, dislocations, twins, voids, cracks).

Understanding the time-dependent, multiscale effects of an element that is difficult to measure is a formidable experimental and computational challenge. Experimental and modelling approaches are required from the atomistic to macro length scales. Recent experimental developments at the atomic length-scale, such as cryo-APT, together with improvements in in-situ testing, both on microscopes and at synchrotron and neutron beamlines, now mean that characterization is more viable. Modelling approaches (including new methods using quantum mechanics, crystal plasticity, evolutionary field models and dislocation dynamics) have recently improved to the point where the time and length scales accessible through experiment and modelling now meet.

Further effort to better relate the role of different microstructural traps on plasticity and fracture modes will allow the creation of hydrogen-induced deformation and fracture maps, similar to conventional plasticity and fracture maps with the additional consideration of the influence of hydrogen. Our fundamental understanding of hydrogen trapping so far has involved only unstressed specimens, and stress is likely to change the energetic landscape that drives trapping behavior. Given that mechanical load is always present, it is necessary in the future to consider the relationship between trap strength and applied stress to better understand the efficacy of hydrogen trapping for withstanding HE.

It may also be possible to develop 'hydrogen trap diagrams', similar to conventional phase diagrams but treat individual traps as thermodynamic entities, considering environmental influences such as temperature, pressure, stress, and hydrogen content related to the structural stability of the trap and the trapping strength for hydrogen, much like defect phase diagrams [445].

The embrittlement of high-strength engineering alloys in hydrogen-containing environments is a longstanding problem. Much has been done. There is much more to do. Despite some progress, the issue is still largely managed in industry by using lower strength, less hydrogen-susceptible alloys, sacrificing efficiencies in design. The dawn of a hydrogen economy completely changes the picture. The hydrogen economy requires the extensive use of engineering alloys for applications from electrolyzers to metallic membranes for gas separation, components for high-pressure hydrogen transport and storage (e.g. pipelines), compressors, liquid hydrogen containers and gas turbines for combustion. This new area brings renewed urgency to mitigating the hydrogen embrittlement challenge.

**Acknowledgements**

Y.-S. C. thanks his family, the Australian Research Council (ARC) Linkage Projects

(LP180100431 and LP210300999), and the 2019 University of Sydney Postdoctoral Fellowship. C.H. thanks China Scholarship Council Postgraduate Research Scholarship (202206120055). P.-Y. L. thanks Taiwan-University of Sydney Scholarship. J. M. C. acknowledges an ARC Future Fellowship (FT180100232). H.-W. Y. thanks Taiwan's Ministry of Science and Technology (NSTC 111-2119-M-002-020-MBK), Advanced Research Center for Green Materials Science and Technology from the Higher Education Sprout Project by the Ministry of Education (112L9006) and the USyd-NTU Partnership Award. E. M.-P. acknowledges financial support from the EPSRC (grants EP/V04902X/1 and EP/V009680/1) and the UKRI Future Leaders Fellows program (grant MR/V024124/1). A. A. acknowledges research support from the Future Fuels CRC.

https://doi.org/10.1016/j.corsci.2012.11.036.

[238] B. Zhang, J. Su, M. Wang, Z. Liu, Z. Yang, M. Militzer, H. Chen, Atomistic insight into hydrogen trapping at MC/BCC-Fe phase boundaries: The role of local atomic environment, Acta Mater. 208 (2021) 116744. https://doi.org/10.1016/J.ACTAMAT.2021.116744.

[239] A. Sadoc, E.H. Majzoub, V.T. Huett, K.F. Kelton, Local structure in hydrogenated Ti–Zr–Ni quasicrystals and approximants, J Alloys Compd. 356–357 (2003) 96–99. https://doi.org/10.1016/S0925-8388(02)01218-5.

[240] J. Radaković, K. Batalović, I. Maarević, J. Belošević-Čavor, Interstitial hydrogen in Laves phases – local electronic structure modifications from first-principles, RSC Adv. 4 (2014) 54769–54774. https://doi.org/10.1039/C4RA09082A.

[241] S.B. Gesari, M.E. Pronsato, A. Visintin, A. Juan, Hydrogen Storage in AB2 Laves Phase (A = Zr, Ti; B = Ni, Mn, Cr, V): Binding Energy and Electronic Structure, Journal of Physical Chemistry C. 114 (2010) 16832–16836. https://doi.org/10.1021/JP106036V.

[242] E.L. Simpson, A.T. Paxton, Effect of applied strain on the interaction between hydrogen atoms and ½ ⟨111⟩ screw dislocations in α-iron, Int J Hydrogen Energy. 45 (2020) 20069–20079. https://doi.org/10.1016/J.IJHYDENE.2020.05.050.

[243] J. Smutna, M.R. Wenman, A.P. Horsfield, P.A. Burr, The bonding of H in Zr under strain, Journal of Nuclear Materials. 573 (2023) 154124. https://doi.org/10.1016/J.JNUCMAT.2022.154124.

[244] S.R. Phillpot, A.C. Antony, L. Shi, M.L. Fullarton, T. Liang, S.B. Sinnott, Y. Zhang, S.B. Biner, Charge Optimized Many Body (COMB) potentials for simulation of nuclear fuel and clad, Comput Mater Sci. 148 (2018) 231–241. https://doi.org/10.1016/J.COMMATSCI.2018.02.041.109